\documentclass[journal]{IEEEtran}
\IEEEoverridecommandlockouts
\usepackage{cite}
\usepackage[ruled,vlined]{algorithm2e}
\usepackage{amsmath,amssymb,amsfonts}
\usepackage{algorithmic}
\usepackage{graphicx}
\usepackage{textcomp}
\usepackage[svgnames]{xcolor}
\usepackage{enumitem}
\usepackage{cancel}
\usepackage{subfigure}
\usepackage{xfrac}
\usepackage[nolist,nohyperlinks]{acronym}

\newtheorem{remark}{\textit{Remark}}
\ifCLASSINFOpdf
 
\else
 
\fi

\def\gil#1{{\color{black}#1}}
\newcommand{\gf}[1]{\textcolor{black}{{#1}}}
\newcommand{\rev}[1]{\textcolor{black}{{#1}}}
\begin{document}
\begin{acronym}
  \acro{IRS}{Intelligent reflecting surface}
  \acro{RIS}{reconfigurable intelligent surface}
   \acro{irs}{intelligent reflecting surface}
  \acro{PARAFAC}{parallel factor analysis}
  \acro{TALS}{trilinear alternating least squares}
  \acro{DF}{decode-and-forward}
  \acro{AF}{amplify-and-forward}
  \acro{CE}{channel estimation}
  \acro{RF}{radio-frequency}
  \acro{THz}{Terahertz communication}
  \acro{EVD}{eigenvalue decomposition}
  \acro{CRB}{Cramér-Rao lower bound}
  \acro{CSI}{channel state information}
  \acro{BS}{base station}
  \acro{MIMO}{multiple-input multiple-output}
   \acro{NMSE}{Normalized Mean Square Error}
   \acro{2G}{Second Generation}
  \acro{3G}{3$^\text{rd}$~Generation}
  \acro{3GPP}{3$^\text{rd}$~Generation Partnership Project}
  \acro{4G}{4$^\text{th}$~Generation}
  \acro{5G}{5$^\text{th}$~Generation}
  \acro{6G}{6$^\text{th}$~generation}
  \acro{E-TALS}{\textit{enhanced} TALS}
  \acro{UT}{user terminal}
  \acro{LS}{least squares}
\end{acronym}

\title{Channel Estimation for Intelligent Reflecting Surface Based MIMO Communication Systems via Tensor Modeling}
\title{Channel Estimation for Intelligent Reflecting Surface Based MIMO Communication Systems: A Tensor Modeling Approach}
\title{Tensor Modeling Approach to Receiver Design for Intelligent Reflecting Surface Assisted MIMO Systems}
\title{A Tensor Approach to Intelligent Reflecting Surface Assisted MIMO Systems: Semi-Blind Joint Channel and Symbol Estimation }
\title{Semi-Blind Joint Channel and Symbol Estimation in \ac{IRS}-Assisted MIMO Systems}
\title{Semi-Blind Joint Channel and Symbol Estimation for \gil{IRS-Assisted} MIMO Systems}

\author{Gilderlan~T.~de~Ara\'{u}jo,
        Andr\'{e}~L.~F.~de~Almeida,~\IEEEmembership{Senior Member,~IEEE,} R\'{e}my Boyer, \textcolor{black}{~\IEEEmembership{Senior Member,~IEEE,}}
        and~G\'{a}bor~Fodor,~\IEEEmembership{Senior Member,~IEEE}
\thanks{Gilderlan T. de Ara\'{u}jo is with Federal Institute of Cear\'{a}, Canind\'{e}, CE, e-mail: gilderlan.tavares@ifce.edu.br}
\thanks{Andr\'{e} L. F. de Almeida is with Wireless Telecommunication Research Group (GTEL), Department of Teleinformatics, Federal University of Cear\'{a}, Fortaleza, CE, e-mail: andre@gtel.ufc.br.}
\thanks{R\'{e}my Boyer is with University of Lille-1, CRIStAL Laboratory, France e-mail: remy.boyer@univ-lille.fr.}
\thanks{Gábor Fodor is with Ericsson Research, 16480 Stockholm, Sweden, and	also with the Division of Decision and Control, KTH Royal Institute of 	Technology, 11428 Stockholm, Sweden, e-mail: gabor.fodor@ericsson.com.}
\thanks{This work was supported by the Ericsson Research, Sweden, and Ericsson Innovation Center, Brazil, under UFC.48
Technical Cooperation Contract Ericsson/UFC. This study was
financed in part by the Coordenação de Aperfeiçoamento de
Pessoal de Nível Superior - Brasil (CAPES)-Finance Code
001, and CAPES/PRINT Proc. 88887.311965/2018-00. Andr\'{e}~L.~F.~de~Almeida acknowledges CNPq for its financial support under the grant 312491/2020-4.}
}

\markboth{Journal of \LaTeX\ Class Files,~Vol.XX, No.~X, XXX}%
{Shell \MakeLowercase{\textit{et al.}}: Bare Demo of IEEEtran.cls for IEEE Journals}

\maketitle

\begin{abstract}
\ac{IRS}  \gil{is} a promising technology \gil{for the  \acl{6G} \gf{of wireless systems}}, realizing the smart radio environment concept. In this paper, we present a novel tensor-based receiver for \ac{IRS}-assisted \gil{\acl{MIMO}} communications capable of jointly estimating the channels and the transmitted data streams in a semi-blind fashion. Assuming a fully passive {\color{black}\ac{IRS} architecture} and introducing a simple space-time coding scheme at the transmitter, \gf{the received signal model can be advantageously built using
the PARATUCK tensor model}, 
which can be seen as a hybrid of \acl{PARAFAC} and Tucker models. Exploiting the algebraic structure of {\color{black}the PARATUCK} tensor model, a semi-blind receiver is derived. The proposed receiver is based on \gf{a} trilinear alternating least squares  method that iteratively estimates the two involved -- \ac{IRS}-\acl{BS} and \acl{UT}-\ac{IRS} -- communication channels and the transmitted symbol matrix. 
We discuss identifiability conditions that ensure the joint semi-blind recovery of the \gil{involved} channel and symbol matrices, and propose a joint design of \gf{the coding and \ac{IRS} reflection matrices} to optimize the receiver performance. For the proposed semi-blind receiver, the derivation of the expected \acl{CRB} is also provided. A numerical performance evaluation of the proposed receiver design corroborates its \gil{superior} performance in terms of the normalized mean square\gil{d} error of the estimated channels \gil{and the achieved symbol error rate.}
\end{abstract}


\IEEEpeerreviewmaketitle
\acresetall 

\renewcommand\baselinestretch{.95}

\acresetall 

\section{Introduction}
\label{introduction}

{\color{black}\ac{IRS} or \acl{RIS} \gf{is a promising technology}  for \ac{6G} wireless systems \cite{Pan_2021}. An \ac{IRS} consists of a \gf{two-dimensional} array of a large number of passive or semi-passive elements, each of which can independently and dynamically tune the desired phase shift and amplitude of the incident radio waves \cite{C_PAN_Edge,Basar_2019}.  
An immediate and simplified application of \acs{IRS} is to overcome the blockage problem between the transmitter and \gf{intended receivers} in wireless networks, which reduces the dead zone\footnote{\color{black}Although a dead zone can be surpassed using relay technology, an \ac{IRS} can be more advantageous in terms of cost since, as opposed to \ac{AF} or \ac{DF} relays that require a dedicated power source, \ac{IRS} does not require power-hungry radio-frequency chains and can also be wirelessly powered by an external RF-based source \cite{Huang2019}. The key differences and similarities between \ac{IRS} and relays are discussed in \cite{Di_HENZO_Relay}.} \cite{Z_PENG_Dead_Zone}. 
\gf{Due to its ability to shape the propagation environment}, \acp{IRS} can be employed in various scenarios to achieve \gf{several other goals as well}, as it is discussed in \cite{DiHenzo_2021}.} \gf{However, deploying \acp{IRS} in wireless communication systems involves a number of challenges, including \ac{CSI} acquisition} \cite{Basar_2019,Survey_NOVO,Big_Tutorial}. 

\gil{Acquiring} \ac{CSI} is an important issue, since the accuracy of the channel estimate has a direct impact on the gains obtained by \ac{IRS}-assisted communications. The difficulty is in part due to the passive nature of the surface and 
the high number of reconfigurable elements.  {\color{black} Recent works have proposed \ac{CE} methods to \ac{IRS}-assisted communications. These works can be classified  according to the \ac{IRS} architecture, system setup and signal processing methodology \cite{zheng_survey}. For example, regarding the \ac{IRS} architecture, it can be assumed that the \ac{IRS} is fully passive}, i.e., it does not have signal processing capabilities and cannot send/process pilot sequences, as it was pointed out in \cite{X_MA_ICC2020}, which investigates \ac{CE} in the context of \ac{IRS}-assisted \acl{THz}. {\color{black}Alternatively, the \ac{IRS} can be semi-passive, where some \ac{IRS} elements are equipped with a few \ac{RF} chains to facilitate the \ac{CE}, as in \cite{Zhang_Semipassive}.}\footnote{\color{black}Note that a fully passive architecture is more challenging, since the estimation of the cascaded channel, TX-\ac{IRS}-Rx, or the individual channels Tx-\ac{IRS} and \ac{IRS}-Rx channels should be carried out at the receiver or transmitter.}
 
 {\color{black} In the system setup category, single and multiuser systems --
 depending on whether the communication links are assisted by a single or multiple \acp{IRS} -- can be distinguished.  As an example,} the authors in \cite{X_Guan_Arxiv2020} and \cite{Zhang_Achor} consider a  multiuser system, and the \ac{CE} solution is based on an anchoring scheme, where two nodes are positioned near the \ac{IRS} in order to aid the \ac{BS}.  Also, references \cite{ZHEN2019,Jawad_2019,C_HU_2019,nadeem2019} propose \ac{CE} strategies, \gf{in which} multi-stage or multi-time scale \gf{estimation techniques are exploited}. 
 A double \ac{IRS}-assisted system is considered in \cite{Zhang_doubleIR}, in which \ac{CE} and a passive beamforming design are investigated.

\gil{Recently, several mechanisms based on deep learning and compressed sensing to acquire \ac{CSI} have been proposed \cite{Jin_2021,Wang_CS}. 
The conventional \ac{LS} \ac{CE} is assumed in \cite{jensen2019optimal}, where a minimum variance unbiased estimator is proposed. \rev{In \cite{Ref_COMPII}, channel training is divided into $I$ blocks. Each block provides a partial channel estimate in the \ac{LS} sense, so that the total channel matrix is accomplished once all blocks are processed.} Paper \cite{Sala_EVD} proposes a matrix factorization method based on \ac{EVD}. \rev{A tensor based solution for IRS-assisted \ac{MIMO} systems is proposed in \cite{Gil_JTSP}. That method relies on a \ac{PARAFAC} modeling of the received signals. It is shown that decoupled estimates of the involved \ac{MIMO} communication channels can be obtained iteratively or in closed form. }
\rev{The works \cite{Alexandropoulos_2020} and \cite{Alexander_SAM2020} also exploit tensor modeling to solve the \ac{CE} task in IRS-assisted downlink multi-user systems.}
} 
The use of tensor methods has been investigated in several previous works in the context of point-to-point \cite{Z_DOU_MIMO2020,Khamidullina_ICASSP2020,DanielCA2019,Sidiropoulos_2019} and cooperative (relay-assisted) \ac{MIMO} systems \cite{F_Du_Acess2020,W_Rui_ICISPC,Yassine_Relay}. The success of tensor-based methods \gf{comes} from the powerful uniqueness properties of tensor decompositions compared to matrix-based ones. Moreover, in wireless communications, tensor-based algorithms efficiently exploit the multi-dimensional nature of the received signals \gf{in the time, space, and frequency domains.} \gf{This multi-dimensional characterization of the received signals leads to more flexible transceiver designs than those offered by conventional matrix-based solutions. Recognizing these benefits of tensor-based algorithms, this paper takes a different approach compared to previous works, and provides joint estimates of the involved communication channels and the transmitted symbols in a semi-blind fashion.}

Assuming a fully passive surface {\color{black}architecture} and introducing a simple space-time coding scheme at the transmitter, we recast the received signal as a PARATUCK tensor model, which can be seen as a hybrid of PARAFAC and Tucker models \cite{Favier2019,PARATUCK_Andre_Elsevier,Harshman_1996,BRO}. Exploiting the algebraic structure of {\color{black}the PARATUCK} tensor model, namely, the different matrix unfoldings of the received signal tensor, a semi-blind receiver based on a \ac{TALS} estimation scheme is proposed. Our receiver design iteratively estimates the two involved (\ac{IRS}-\ac{BS} and \ac{UT}-\ac{IRS}) communication channels and the symbol matrix. Moreover, by resorting to the identifiability results of the PARATUCK tensor model, we derive useful system design recommendations that ensure the joint semi-blind recovery of the \gil{involved} channel and the symbol matrices. In particular, we propose a joint design of the coding matrix and the \ac{IRS} reflection matrix to optimize the receiver performance. We also present an extension of the proposed receiver algorithm to a scenario, in which the direct link is available. \gf{In this scenario, an initial estimation of the transmitted symbol matrix obtained from the direct link is used as a warm start to enhance the semi-blind joint channel and symbol estimation \textit{via} the \ac{IRS}-assisted link.} Finally, we provide expressions for the expected \ac{CRB} for the proposed semi-blind receiver. 

In the following, we summarize the main contributions of this work.
{\color{black} 
\begin{itemize}
    
    \item \gil{We present a novel tensor-based semi-blind receiver algorithm for \ac{IRS}-assisted \ac{MIMO} systems that avoids the \emph{a priori} training phase. The proposed algorithm iteratively estimates the two involved channel matrices as well as the symbol matrix by means of a \ac{TALS} 
    algorithm, which exploits a PARATUCK tensor model for the received signals.}
    
    \item We extend the proposed semi-blind receiver to a scenario, in which the direct link between transmitter and receiver is available. In this case, the receiver processing has two stages. In the first one, an initial semi-blind estimate of the transmitted symbols obtained \textit{via} the direct link is used as a warm start of the joint channel and symbol estimation \textit{via} the \ac{IRS}-assisted link.
    
   \item \rev{We derive conditions for the joint channel and symbol identifiability, and discuss the design of the {\color{black}coding matrix} and the \ac{IRS} phase shift matrix. A joint design is proposed to {\color{black}improve} the receiver performance.}
   
    \item \gil{ \rev{We investigate the impact of the direct channel estimation on the accuracy of the \ac{IRS}-assisted channel estimation, showing that a good estimate of the direct channel in the first stage improves the performance of the joint channel and symbol estimate in the second stage.}}
    
    \item We derive the expected \ac{CRB} for the proposed TALS-PARATUCK semi-blind receivers, allowing to study its performance analytically.
\noindent     
\end{itemize}
}

\ 

\vspace{-2ex}
\noindent \textit{Notation and properties}: Matrices are represented with boldface capital letters ($\mathbf{A})$, and vectors are denoted by boldface lowercase letters ($\mathbf{a})$. Tensors are symbolized by calligraphic letters $(\mathcal{A})$. Transpose and pseudo-inverse of a matrix $\mathbf{A}$ are denoted as $\mathbf{A}^{\textrm{T}}$ and $\mathbf{A}^\dagger$, respectively.  {\color{black}$D_i(\mathbf{A})$ is a diagonal matrix holding the $i$th row of $\mathbf{A}$ on its main diagonal}. The operator $\textrm{diag}(\mathbf{a})$ forms a diagonal matrix out of its vector argument, while $\ast$, $\circ$, $\diamond$, $\odot$ and $\otimes$ denote the conjugate,  outer product, Khatri Rao, Hadamard and Kronecker products, respectively. $\mathbf{I}_N$ denotes the $N \times N$ identity matrix. The operator $\textrm{vec}(\cdot)$ vectorizes an $I \times J$ matrix argument, while $\textrm{unvec}_{I \times J}(\cdot)$ does the opposite operation. Moreover, $\textrm{vecd}(.)$ forms a vector out of the diagonal of its matrix argument. ¨The $n$-mode product between a tensor $\mathcal{Y} \in \mathbb{C}^{I \times J \ldots \times K}$ and a matrix $\mathbf{A} \in \mathbb{C}^{I \times R}$ is denoted as $\mathcal{A}\times_n\mathbf{B}$, for $1 \leq n \leq N $. An identity $N$-way tensor of dimension $R\times R \cdots \times R$ is denoted as $\mathcal{I}_{N,R}$. Moreover, {\color{black}$\mathbf{A}_{i\bullet}$ and $\mathbf{A}_{\bullet j}$ denotes the $i$-th row and $j$-th column of the matrix $\mathbf{A}$, respectively}. The operator $\left\lceil{x}\right\rceil$ rounds its fractional argument up to the nearest integer. In this paper, we make use of the following identities:
\begin{equation}
(\mathbf{A} \diamond \mathbf{B})^{\textrm{H}}(\mathbf{C} \diamond \mathbf{D}) = (\mathbf{A}^{\textrm{H}}\mathbf{C}) \odot  (\mathbf{B}^{\textrm{H}}\mathbf{D}).
\label{Propertie Hadmard x Khatri}
\end{equation}
\begin{equation}
\textrm{vec}(\mathbf{A\mathbf{B}\mathbf{C}}) = (\mathbf{C}^{\textrm{T}} \otimes \mathbf{A})\textrm{vec}(\mathbf{B}).
\label{Propertie Vec General}
\end{equation}
\begin{equation}
\textrm{diag}(\mathbf{a})\mathbf{b} = \textrm{diag}(\mathbf{b})\mathbf{a}.
\label{Propertie diag(a)b}
\end{equation}
If $\mathbf{B}$ is a diagonal matrix, we have:
\begin{equation}
\textrm{vec}(\mathbf{A\mathbf{B}\mathbf{C}}) = (\mathbf{C}^{\textrm{T}} \diamond \mathbf{A})\textrm{vecd}(\mathbf{B}).
\label{Propertie Vec restrict}
\end{equation}

{\color{black}\section{Tensor preliminaries}
\label{Sec: Tensor preliminaries}
In this section, we provide a brief overview on two tensor decompositions that are of interest to this work, namely the PARAFAC and PARATUCK decompositions. They will be exploited in the formulation of the proposed receivers. In order to keep the presentation concise, the focus is on the key definitions and expressions used to represent these two tensor decompositions. 

\vspace{-2ex}
\gil{\subsection{PARAFAC decomposition}
\label{Overview PARAFAC}
The PARAFAC decomposition, also known as the canonical polyadic  decomposition (CPD), is the most popular tensor decomposition, which expresses a tensor as a sum of a minimum number of rank-one tensors \cite{Kolda_TDA,Hitchcock,Carrol_AND_CHANG,Harshman70}. For a third-order tensor $\mathcal{X} \in \mathbb{C}^{I \times J \times K}$, its scalar form and frontal slice representation is given as
\begin{equation}
x_{i,j,k} = \sum_{r = 1}^{R}a_{i,r}b_{j,r}c_{k,r},
\label{CPD:ScalarNotation}
\end{equation}
and 
\vspace{-1ex}
\begin{equation}
\mathbf{X}[k] = \mathbf{A}D_k(\mathbf{C})\mathbf{B}^{\textrm{T}}\, \, \in \mathbb{C}^{I \times J}, \label{slice_3}
\end{equation}
respectively, where $x_{i,j,k}$ denotes the $(i,j,k)$-th entry of the tensor $\mathcal{X} \in \mathbb{C}^{I \times J \times K}$ and $\mathbf{X}[k]$ is the $k$-th frontal slice (a.k.a. as $3$-mode slice) of the tensor $\mathcal{X}$, for $k=1,\ldots, K$. The scalars $a_{i,r}$, $b_{j,r}$ and $c_{k,r}$ are corresponding entries of the three factor matrices $\mathbf{A}$, $\mathbf{B}$, and $\mathbf{C}$, while $R$ denotes the \textit{rank} of the tensor $\mathcal{X}$. 
where $\mathbf{X}[k]$ is the $k$-th frontal slice (a.k.a. as 3-mode slice) of the tensor $\mathcal{X}$, $k=1,\ldots, K$. The PARAFAC decomposition is powerful due to its essential factor identification uniqueness property, which has its roots on the concept of the Kruskal rank (k-rank). Further details can be found in \cite{KRUSKAL197795, STEGEMAN2007540}.}

\vspace{-2ex}
\subsection{PARATUCK decomposition}
\label{Overview tensor}
The PARATUCK decomposition \cite{Harshman_1996,Favier_EUSIPICO} is a hybrid tensor decomposition that combines the Tucker \cite{Tucker1966} and the PARAFAC decompositions. It enjoys the powerful uniqueness properties of the PARAFAC model, while offering a more flexible structure by \rev{allowing controlled interactions among its factor matrices.} Its scalar form and frontal slice representations are given as 
\begin{equation}
x_{i,j,k} = \sum_{r_1 = 1}^{R_1}\sum_{r_2 = 2}^{R_2}a_{i,r_1}b_{j,r_2}\omega_{r_1,r_2}c_{k,r_1}^{A}c_{k,r_2}^{B},
\label{PT2:scalarNotation}
\end{equation}
and 
\begin{equation}
\mathbf{X}[k] = \mathbf{A}D_k(\mathbf{C}^A)\boldsymbol{\Omega}D_k(\mathbf{C}^B)\mathbf{B}^{\textrm{T}},
\label{PT2:SliceNotation}
\end{equation}

respectively, where $a_{i,r_1}$, $b_{j,r_2}$, $\omega_{r_1,r_2}$, $c_{k,r_1}^{A}$ and $c_{k,r_2}^{B}$ are the elements of the matrices $\mathbf{A} \in \mathbb{C}^{I \times R_1}$, $\mathbf{B} \in \times \mathbb{C}^{J \times R_2}$,  $\boldsymbol{\Omega} \in \mathbb{C} ^{R_1 \times R_2}$, $\mathbf{C}^{A} \in \mathbb{C}^{K \times R_1}$ and $\mathbf{C}^{B} \in \mathbb{C}^{K \times R_2}$, respectively. $\mathbf{A}$ and $\mathbf{B}$ are referred to as the \emph{factors matrices}, $\mathbf{C}^{A}$ and $\mathbf{C}^{B}$ are the \emph{interactions matrices}, while $\boldsymbol{\Omega}$ is the \emph{core matrix}, whose $(r_1,r_2)$-th entry defines the level of interaction between the $r_1$-th column of $\mathbf{A}$ and the $r_2$-th column of $\mathbf{B}$. 
}

\vspace{-2ex}
\section{System Model}
\label{Sec: System_Model}
\begin{figure}[!t]
	\centering\includegraphics[scale=0.65]{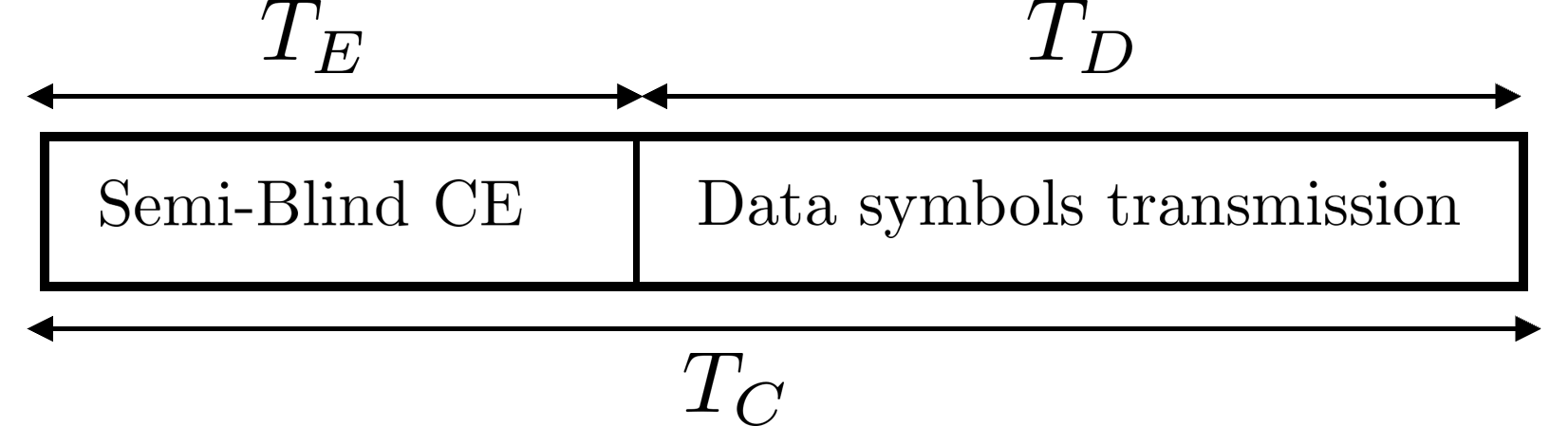}
	\caption{Transmission time structure.}
	\label{fig:Trans_Protocol}
\end{figure}

Let us consider  a \ac{MIMO} communication system assisted by an \ac{IRS}, where the \ac{BS} and \ac{UT} have arrays of $M$ and $L$ antennas, respectively, while the \ac{IRS} is composed of $N$ elements, which can be individually adjusted/configured to generate phase shifts. \gil{We assume a two-stage structured time-domain transmission, as illustrated in Fig. \ref{fig:Trans_Protocol}, where the estimation time  $T_E$} is split into $K$ blocks, and each block has $T$ symbol periods each, with \gil{$T_E = KT \leq T_C$}. We focus on the uplink scenario, where a multi-antenna \ac{UT} encodes $L$ independent data streams that are received at the \ac{BS} with the assistance of an \ac{IRS}, and possibly reach the \ac{BS} directly (if the direct link is available).\footnote{\color{black}Although the uplink case is assumed here, the signal model and the algorithms proposed in this paper are equally applicable to the downlink case by just inverting the roles of the transmitter and the receiver.} 

In the general case (when the direct link available), the discrete-time baseband received signal vector during the $t$-th symbol period of the $k$-th block is given by
\gil{\begin{equation}
  \mathbf{y}[k,t] = \underbrace{\mathbf{H}^{\textrm{(D)}}\mathbf{x}[k,t]}_{\textrm{Direct link}} +  \underbrace{\mathbf{H}\textrm{diag}(\mathbf{s}[k,t])\mathbf{G}\mathbf{x}[k,t]}_{\textrm{\ac{IRS}-assisted link}} + \mathbf{b}[k,t]\, ,
   \label{Eq:Receiver_Signal}
\end{equation}
}
where $\mathbf{H}^{\textrm{(D)}} \in \mathbb{C}^{M \times L}$ is the direct channel matrix between the \ac{UT} and the \ac{BS}, whereas $\mathbf{H} \in \mathbb{C}^{M \times N}$ and $\mathbf{G} \in \mathbb{C}^{N \times L}$ denotes the \ac{IRS}-\ac{BS} and \ac{UT}-\ac{IRS} channel matrices, respectively. {\color{black}The \ac{UT} employs a space-time encoding scheme that ``diagonally'' encodes the input symbols such that $\mathbf{x}[k,t] = \textrm{diag}(\mathbf{w}[k,t])\mathbf{x}[t] \in \mathbb{C}^{L \times 1}$} contains the encoded symbol vector, $\mathbf{x}[t]$,  transmitted during the $t$-th symbol period of the $k$-th block.  The vector $\mathbf{s}[k,t] \in \mathbb{C}^{N \times 1}$ collects the \ac{IRS} phase shifts, and $\mathbf{b}[k,t] \in \mathbb{C}^{M \times 1}$ is the corresponding additive white Gaussian noise vector. {\color{black}We assume that the phase shift vector $\mathbf{s} \in \mathbb{C}^{N \times 1}$ and the coding vector $\mathbf{w} \in  \mathbb{C}^{L \times 1}$ are constant during the $T$ time slots of the $k$-th block, and vary from block to block, which means that $\mathbf{s}[k,t] = \mathbf{s}[k] $ and $ \mathbf{w}[k,t] = \mathbf{w}[k]$, for $1\leq t\leq T$. 

With these assumptions, collecting the received signals during the $T$ time slots of each block yields
\begin{equation}
  \mathbf{Y}[k] = \mathbf{H}^{\textrm{(D)}}D_{k}(\mathbf{W})\mathbf{X}^{\textrm{T}} + \mathbf{H}D_{k}(\mathbf{S})\mathbf{G}D_{k}(\mathbf{W})\mathbf{X}^{\textrm{T}} + \mathbf{B}[k]\, ,
   \label{EQ:General_PT2} 
\end{equation}
where $\mathbf{Y}[k]\doteq [\mathbf{y}[k,1], \ldots, \mathbf{y}[k,T]] \in \mathbb{C}^{M \times T}$ collects the received signal vectors during the $t= 1, \ldots, T$ time slots of the $k$-th block.}  {\color{black}In this paper,} we consider two possible scenarios. In the first one, the direct link is assumed to be weak, or unavailable. In the second one, both the \ac{IRS}-assisted link and the direct link are exploited.

\begin{figure}[!t]
	\centering\includegraphics[scale=0.45]{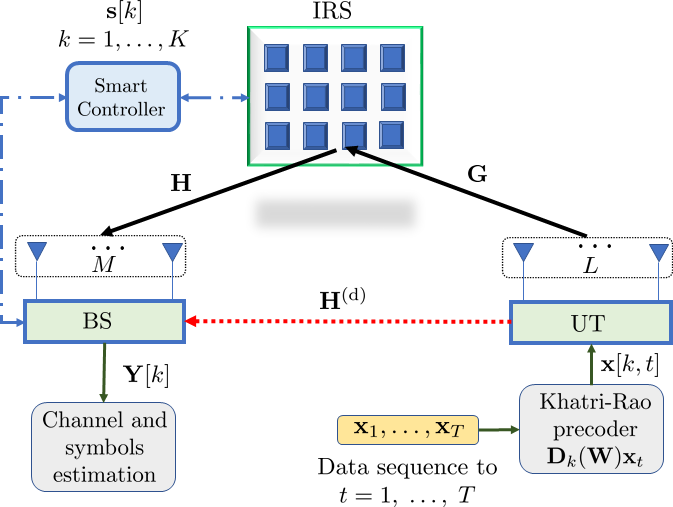}
	\caption{Uplink \ac{IRS}-assisted \ac{MIMO} system diagram.}
	\label{fig:LIS}
\end{figure}

\begin{figure}[!t]
    \includegraphics[scale = 0.55]{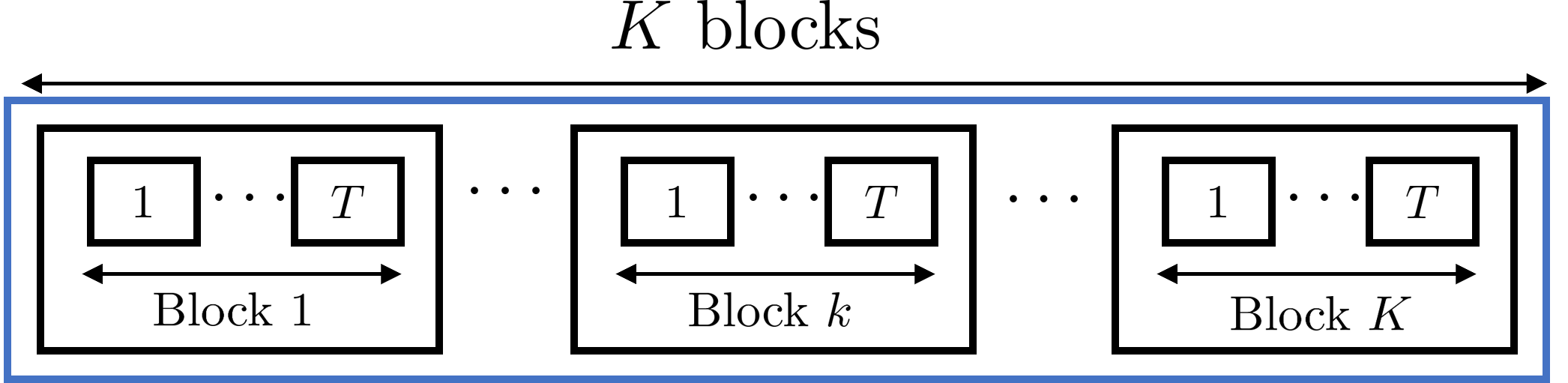}
	\caption{\rev{Time protocol when the direct \ac{BS}-\ac{UT} link is not available.}}
	\label{fig:Time pattern without direct link}
\end{figure}

{\color{black}Regarding the channel model, no particular assumption is made in this paper. We can assume that the channel matrices follow an i.i.d Rayleigh fading model, or, alternatively, are described by a geometrical model with a few specular paths. For instance, we can assume that the \ac{UT}-\ac{IRS} and \ac{IRS}-\ac{BS} links are subject to \gf{low scattering propagation}, such that $\mathbf{H} = \mathbf{A}_{\textrm{IRS}}\textrm{diag}(\boldsymbol{\beta})\mathbf{A}_{\textrm{BS}}^{\textrm{H}}$, and $\mathbf{G} = \mathbf{B}_{\textrm{UT}}\textrm{diag}(\boldsymbol{\gamma})\mathbf{B}_{\textrm{IRS}}^{\textrm{H}}$, where $ \mathbf{A}_{\textrm{BS}} \in \mathbb{C}^{M \times R_1}$, $\mathbf{A}_{\textrm{IRS}} \in \mathbb{C}^{N \times R_1}$, $\mathbf{B}_{\textrm{UT}} \in \mathbb{C}^{L \times R_2}$ and $\mathbf{B}_{\textrm{IRS}} \in \mathbb{C}^{N \times R_2}$ are the array response matrices, and the vectors $\boldsymbol{\beta}$ and $\boldsymbol{\gamma}$ hold the complex amplitude coefficients of the \ac{IRS}-\ac{BS} and \ac{UT}-\ac{IRS} channels, respectively, while $R_1$ and $R_2$ denote the number of clusters between \ac{IRS}-\ac{BS} and \ac{UT}-\ac{IRS}, respectively \cite{Heath2016}.}

\section{Semi-blind receiver without the direct link}

{\color{black}Considering the first scenario, when the direct link is weak or unavailable, the received signal in (\ref{EQ:General_PT2}) reduces to}
\begin{equation}
  \mathbf{Y}[k]=  \mathbf{H}D_k(\mathbf{S})\mathbf{G}D_k(\mathbf{W})\mathbf{X}^{\textrm{T}} + \mathbf{B}[k],
   \label{Signal:Paratuck2} 
\end{equation}

where $\mathbf{S}\doteq [\mathbf{s}[1], \ldots, \mathbf{s}[K]]^{\textrm{T}}  \in \mathbb{C}^{K \times N}$, $\mathbf{W}\doteq [\mathbf{w}[1], \ldots, \mathbf{w}[K]]^{\textrm{T}}  \in \mathbb{C}^{K \times L}$ are the phase shift matrix and {\color{black}coding} matrix, {\color{black}respectively,} and $\mathbf{X}\doteq [\mathbf{x}[1], \ldots, \mathbf{x}[T]]^{\textrm{T}} \in \mathbb{C}^{T \times L}$ is the transmitted symbol matrix. 
Note that the useful (signal) part of the received signal during the $k$-th block can be identified as the $k$-th matrix slice of a \textit{received signal tensor} $\mathcal{Y} \in \mathbb{C}^{M \times T \times K}$ that satisfies a PARATUCK decomposition \cite{PARATUCK_Andre_Elsevier,Harshman_1996}, {\color{black} where } the scalar form of the noiseless received signal tensor $\overline{\mathcal{Y}}$ can be expressed as
\begin{equation}\label{eq:paratuck_scalar}
\overline{y}_{m,t,k} = \sum_{n = 1}^{N}\sum_{l = 1}^{L}g_{n,l}x_{t,l}h_{m,n}s_{k,n}w_{k,l}.
\end{equation}
Note that the interaction matrices of the PARATUCK model correspond to the matrices $\mathbf{S}$ and $\mathbf{W}$ that collect, respectively, the phase shifts (introduced by the \ac{IRS}) and the {\color{black}coding} coefficients (applied at the transmitter), which are fixed and known at the receiver. {\color{black}To summarize, comparing equations  (\ref{PT2:SliceNotation}) and (\ref{Signal:Paratuck2}), the following correspondence can be established
\begin{eqnarray}
\left( \mathbf{A}, \mathbf{B}, \mathbf{R}, \mathbf{C}^\mathbf{A}, \mathbf{C}^\mathbf{B}\right) \leftrightarrow (\mathbf{H}, \mathbf{X}, \mathbf{G}, \mathbf{S}, \mathbf{W}).
\label{EQ:PARATUCK_correspondece}
\end{eqnarray}
}
The algebraic properties of this PARATUCK tensor signal model will be exploited to formulate our semi-blind receiver for joint channel and symbol estimation. {\color{black} A more complete scenario, in which the direct link is available, as indicated in (\ref{EQ:General_PT2}), will be discussed later.}

\begin{remark}
The \ac{IRS}-assisted channel is usually represented in an equivalent form in which the channels $\mathbf{G}$ and $\mathbf{H}$ are linked by a Khatri-Rao product. This link can be seen by defining the channel parameter vector $\boldsymbol{\theta} = \textrm{vec}(\mathbf{G}^{\textrm{T}} \diamond \mathbf{H})$. In general, $\boldsymbol{\theta}$ can be directly estimated in the LS sense from the received signal, or constructed once the individual estimates of $\mathbf{G}$ and  $\mathbf{H}$ are obtained. In this paper, we adopt the second approach.
\end{remark}

\label{Sec: PARTAUCK Receiver Fomulation}
Our goal is to jointly estimate all the \ac{UT}-\ac{IRS} channel $\mathbf{G} \in \mathbb{C}^{N \times L}$, the \ac{IRS}-\ac{BS} channel $\mathbf{H} \in \mathbb{C}^{M \times N}$, and the symbol matrix $\mathbf{X} \in \mathbb{C}^{T \times L}$ by exploiting the tensor structure of the received signal model (\ref{Signal:Paratuck2}). We start by stating the following optimization problem:
\begin{equation}
 \underset{\mathbf{H},\mathbf{G},\mathbf{X}}{\textrm{min}} \,\,\sum\limits_{k=1}^K\Big\| \mathbf{Y}[k]  -  \mathbf{H}D_k(\mathbf{S})\mathbf{G}D_k(\mathbf{W})\mathbf{X}^{\textrm{T}} \Big\|_F.
   \label{cost_func}
\end{equation}
Clearly, this problem is highly nonlinear, since it involves multiple products of the unknown variables, represented by the matrices $\mathbf{H}, \mathbf{G}$, and $\mathbf{X}$. However, we take a simpler route to solve the above problem by capitalizing on the multi-linear nature of the received signal and exploiting the PARATUCK tensor model structure \cite{PARATUCK_Andre_Elsevier,Harshman_1996}. By operating on each one of the three different matrix unfoldings of this tensor model, we derive the key equations for conditionally minimizing the cost function (\ref{cost_func}) with respect to each unknown matrix in the least squares (LS) sense, while assuming that the remaining quantities are fixed. To simplify the presentation, we temporarily omit the noise term during the development of the main steps. 

\subsection{Estimation of the \ac{IRS}-\ac{BS} channel}
Let us first consider the estimation of the \ac{IRS}-\ac{BS} channel matrix. Starting from the frontal slice representation in (\ref{Signal:Paratuck2}), and stacking column-wise the $K$ matrix slices $\{\mathbf{Y}[k]\}$, $k=1, \ldots, K$, we get
\begin{equation}
\begin{aligned}
\mathbf{Y}_1 &\doteq [\mathbf{Y}[1], \ldots, \mathbf{Y}[K]]=\mathbf{H}\mathbf{F}^{\textrm{T}} + \mathbf{B}_1\, \, \in \mathbb{C}^{M \times TK},
\end{aligned}
\label{Estimate_G}  
\end{equation}
which corresponds to the 1-mode unfolding of the received signal tensor in (\ref{eq:paratuck_scalar}),
where 
\begin{equation}
    \mathbf{F} \doteq \left[
    \begin{array}{c}
       \mathbf{X}D_1(\mathbf{W})\mathbf{G}^{\textrm{T}}D_1(\mathbf{S})\\
       \vdots \\
       \mathbf{X}D_K(\mathbf{W})\mathbf{G}^{\textrm{T}}D_K(\mathbf{S})
    \end{array}
    \right] \;\; \in \; \mathbb{C}^{TK \times N},\label{eq:Fmatrix}
\end{equation}
and $\mathbf{B}_1$ is the corresponding 1-mode unfolding of the additive noise tensor. The estimation of $\mathbf{H}$ can be obtained by solving the following LS problem
\begin{equation}
\Hat{\mathbf{H}} = \underset{\mathbf{H}}{\arg\min} \,\, \left\|\mathbf{Y}_{1} -\mathbf{H}\mathbf{F}^{\textrm{T}}\right\|_F^2,
\label{func costG_PT2}
\end{equation}
the solution of which is given by
\begin{equation}
    \Hat{\mathbf{H}} = \mathbf{Y}_1\left( \mathbf{F}^{\textrm{T}}\right)^{\dagger}.
    \label{Estimate_G_LS}  
\end{equation}

\subsection{\ac{UT}-\ac{IRS} channel estimation}
To derive the update equation for the estimation of the \ac{UT}-\ac{IRS} channel matrix $\mathbf{G}$, let us apply the $\textrm{vec}(.)$ operator to (\ref{Signal:Paratuck2}), which gives
\begin{equation}
\begin{aligned}
&\textrm{vec}\left(\mathbf{Y}[k]\right)  = (\mathbf{X} \otimes \mathbf{H})\textrm{vec}\left(D_k(\mathbf{S})\mathbf{G}D_k(\mathbf{W})\right)\\
& = (\mathbf{X} \otimes \mathbf{H})\left(D_k(\mathbf{W}) \otimes D_k(\mathbf{S})\right)\textrm{vec}(\textbf{G}) + \textrm{vec}\left(\mathbf{B}[k]\right)\, ,
\end{aligned}
\label{VEC_Equation Key 35}
\end{equation}
where we have applied property (\ref{Propertie Vec General}) twice. Now, applying property (\ref{Propertie diag(a)b}) to  (\ref{VEC_Equation Key 35}) yields
\begin{equation}
\textrm{vec}\left(\mathbf{Y}[k]\right) = (\mathbf{X} \otimes \mathbf{H})\textrm{diag}(\textrm{vec}(\mathbf{G}))\left(\mathbf{W}_{k\bullet}^{\textrm{T}} \otimes \mathbf{S}_{k\bullet}^{\textrm{T}}\right) + \textrm{vec}\left(\mathbf{B}[k]\right), 
\label{Concatenating VEC(35)}
\end{equation}
where we have used the fact that $\left(D_k(\mathbf{W}) \otimes D_k(\mathbf{S})\right)$ is actually $\textrm{diag}\left(\mathbf{W}_{k\bullet}^{\textrm{T}} \otimes \mathbf{S}_{k\bullet}^{\textrm{T}}\right)$. By stacking column-wise $\textrm{vec}\left(\mathbf{Y}[1]\right), \ldots, \textrm{vec}\left(\mathbf{Y}[K]\right)$, and using (\ref{Concatenating VEC(35)}), we can obtain the 3-mode unfolding of the received signal tensor as follows
\begin{equation}
\begin{aligned}
\mathbf{Y}_3 &\doteq \left[\textrm{vec}\left(\mathbf{Y}[1]\right), \ldots, \textrm{vec}\left(\mathbf{Y}[K]\right)\right]\\
&= (\mathbf{X} \otimes \mathbf{H})\textrm{diag}(\textrm{vec}(\mathbf{G}))\mathbf{\Psi}  + \mathbf{B}_3\, ,\, \in \,  \mathbb{C}^{TM \times K}
\end{aligned}
\label{VEC to ESTIMATE H}
\end{equation}
where 
\begin{equation}
\begin{aligned}
\mathbf{\Psi} & \doteq \left[\mathbf{W}_{1\bullet}^{\textrm{T}} \otimes \mathbf{S}_{1\bullet}^{\textrm{T}}, \ldots, \mathbf{W}_{K\bullet}^{\textrm{T}} \otimes \mathbf{S}_{K\bullet}^{\textrm{T}}\right]\\
& =\mathbf{W}^{\textrm{T}} \diamond \mathbf{S}^{\textrm{T}} \in \mathbb{C}^{LN \times K}. \label{eq:defPsi}
\end{aligned}
\end{equation}
Finally, vectorizing (\ref{VEC to ESTIMATE H}) and applying  property (\ref{Propertie Vec General}) yields
\begin{equation}
\begin{aligned}
\textrm{vec}\left(\mathbf{Y}_3\right) &= \left[\mathbf{\Psi}^\textrm{T} \diamond (\mathbf{X} \otimes \mathbf{H})\right]\textrm{vec}(\mathbf{G}) + \textrm{vec}\left(\mathbf{B}_3\right).
\end{aligned}
\label{Estimate_H}
\end{equation}
Thus, an estimate of $\mathbf{G}$ in the LS sense can be obtained by solving the following problem
\begin{equation}
\Hat{\mathbf{G}} = \underset{\mathbf{G}}{\arg\min} \,\, \left\|\textrm{vec}\left(\mathbf{Y}_3\right) -\left[\mathbf{\Psi}^\textrm{T} \diamond (\mathbf{X} \otimes \mathbf{H})\right]^{\dagger} \textrm{vec}(\mathbf{G})\right\|_F^2,
\label{func costH_PT2}
\end{equation}
the solution of which is given by
\begin{equation}
     \Hat{\mathbf{G}} =\textrm{unvec}_{N \times L}\Big(\left[\mathbf{\Psi}^\textrm{T} \diamond (\mathbf{X} \otimes \mathbf{H})\right]^{\dagger} \textrm{vec}(\mathbf{Y}_3)\Big).
    \label{H_hat}
\end{equation}
\begin{remark}
The step 2 of Algorithm 1, which is concerned with the estimation of the \ac{UT}-\ac{IRS} channel matrix, can be simplified by assuming that $\mathbf{\Psi}$ defined in (\ref{eq:defPsi}) is an $LN \times K$ semi-unitary matrix satisfying $\mathbf{\Psi}^{\ast}\mathbf{\Psi}^{\textrm{T}}=K\mathbf{I}_{LN}$ (this choice is discussed in Appendix \ref{Appendix B}). In this case, it can be shown  that the LS estimation step (\ref{H_hat}) simplifies to
\begin{equation}
    \textrm{vec}(\mathbf{G}) = (1/K)\cdot\boldsymbol{\Sigma}_{\mathbf{Q}}^{-1}(\mathbf{\Psi}^{\textrm{T}} \diamond \mathbf{Q})^{\textrm{H}}\textrm{vec}\left( \mathbf{Y}_3 \right)\label{H_hat_2},
\end{equation}
where $\mathbf{Q} \doteq [\mathbf{q}_1, \ldots, \mathbf{q}_{LN}]= \mathbf{X} \otimes \mathbf{H} \in \mathbb{C}^{TM \times LN}$, and 
\begin{equation}
   \boldsymbol{\Sigma}_{\mathbf{Q}} \doteq \left[\begin{array}{ccc}
      \|\mathbf{q}_1\|^2   &  &\\
         & \ddots &\\
         &   &  \|\mathbf{q}_{LN}\|^2
    \end{array}\right].
    \label{EQ:Sigma_Q}
\end{equation}
In addition to the complexity reduction, our numerical experiments have shown that the semi-unitary design for $\mathbf{\Psi}$ also improves the convergence speed of Algorithm 1. On the other hand, this condition requires $K \geq LN$. It is worth noting, however, that although advantageous from a performance/complexity viewpoint, the semi-unitary condition is not necessary.
\end{remark} 

\IncMargin{1em}
\begin{algorithm}[!t]
\footnotesize{
	\DontPrintSemicolon
	\SetKwData{Left}{left}\SetKwData{This}{this}\SetKwData{Up}{up}
	\SetKwFunction{Union}{Union}\SetKwFunction{FindCompress}{FindCompress}
	\SetKwInOut{Input}{input}\SetKwInOut{Output}{output}
	\textbf{Procedure}\\
	\Input{$i = 0$; \textrm{Initialize} $\Hat{\mathbf{G}}_{(i=0)}$ \textit{and} $\Hat{\mathbf{X}}_{(i=0)}$}
	\Output{$\Hat{\mathbf{H}}$, $\Hat{\mathbf{G}}$ and $\Hat{\mathbf{X}}$}
	\BlankLine

	\Begin{
		$i = 1 ;$\;
		\While{$\|e(i) - e(i-1)\| \geq \delta$}{
			\vspace{2ex}
			\begin{enumerate}
				\item [1.] \textrm{Using} $\Hat{\mathbf{G}}_{(i-1)}$ \textrm{and} $\Hat{\mathbf{X}}_{(i-1)},$ \textrm{compute}\\ $\Hat{\mathbf{F}}_{(i-1)}$ \textrm{from} (\ref{eq:Fmatrix}) \textrm{and find \\ a least squares estimate of} $\mathbf{H}$:\\
				\vspace{1ex}
				$\Hat{\mathbf{H}}_{(i)} = \mathbf{Y}_1\left( \Hat{\mathbf{F}}^{\textrm{T}}_{(i-1)}\right)^{\dagger}$
				\vspace{2ex}
				\item[2. ] \textrm{Using} $\Hat{\mathbf{H}}_{(i)}$ and $\Hat{\mathbf{X}}_{(i-1)}$, find\\ 
				\textrm{a least squares estimate of $\mathbf{G}$}:\\
				\vspace{1ex}
				$\textrm{vec}(\Hat{\mathbf{G}}_{(i)})= \left[\mathbf{\Psi}^\textrm{T} \diamond (\mathbf{X}_{(i-1)} \otimes \mathbf{H}_{(i)})\right]^{\dagger} \textrm{vec}(\mathbf{Y}_3)$
			\vspace{2ex}
				\item[3.] \textrm{Using} $\Hat{\mathbf{G}}_{(i)}$ and $\Hat{\mathbf{H}}_{(i)},$ \textrm{compute}\\ $\Hat{\mathbf{E}}_{(i)}$ \textrm{from} (\ref{eq:Ematrix}) \textrm{and find \\ a least squares estimate of} $\mathbf{X}$:\\
				\vspace{1ex}
				$\Hat{\mathbf{X}}_{(i)} = \mathbf{Y}_2\left( \Hat{\mathbf{E}}^{\textrm{T}}_{(i)}\right)^{\dagger}$
				\vspace{2ex}
				\item[4:] $i \leftarrow i+1$
				\item[5:]\textrm{Repeat steps} $1$ to $4$ \textrm{until convergence.}
			\end{enumerate}
			\textbf{end}
		}
	\textbf{end}	
	}
	\caption{\ac{TALS}}\label{PsCode_PARATUCK}
	\label{Algorithm:TALS}
	}
\end{algorithm}\DecMargin{0.8em}

\subsection{Symbol estimation}
The final step of our semi-blind receiver estimates the transmitted symbol matrix. To this end, we start from (\ref{Signal:Paratuck2}), and stack column-wise the matrix slices $\mathbf{Y}[1], \ldots, \mathbf{Y}[K]$, to get
\begin{equation}
\begin{aligned}
\mathbf{Y}_2 &\doteq [\mathbf{Y}[1]^\textrm{T}, \ldots, \mathbf{Y}[K]^\textrm{T}]=\mathbf{X}\mathbf{E}^{\textrm{T}} \, \, \in \mathbb{C}^{T \times MK},
\end{aligned}
\label{Estimate_X}
\end{equation}
which corresponds to the 2-mode unfolding of the received signal tensor in (\ref{eq:paratuck_scalar}), where
\begin{equation}
\mathbf{E} \doteq \left[\begin{array}{c}
\mathbf{H}D_1(\mathbf{S})\mathbf{G}D_1(\mathbf{W})\\
\vdots\\
\mathbf{H}D_K(\mathbf{S})\mathbf{G}D_K(\mathbf{W})
\end{array}\right]\in \mathbb{C}^{MK \times L}.\label{eq:Ematrix}
\end{equation}
Adding the noise term, we have $\mathbf{Y}_2= \mathbf{X}\mathbf{E}^{\textrm{T}} + \mathbf{B}_2$. The LS estimate of $\mathbf{X}$ is then obtained by solving
\begin{equation}
\Hat{\mathbf{X}} = \underset{\mathbf{X}}{\arg\min} \,\, \left\|\mathbf{Y}_{2} -\mathbf{X}\mathbf{E}^{\textrm{T}}\right\|_F^2,
\label{func costX_PT2}
\end{equation}
the solution of which is given by
\begin{equation}
    \Hat{\mathbf{X}} = \mathbf{Y}_2\left( \mathbf{E}^{\textrm{T}}\right)^{\dagger}.
        \label{X_hat}
\end{equation}

The proposed semi-blind receiver makes use of  (\ref{Estimate_G_LS}), (\ref{H_hat_2}) and (\ref{X_hat}) to obtain the estimates of channel matrices $\mathbf{G}$ and $\mathbf{H}$, and the symbol $\mathbf{X}$ via a trilinear alternating least squares based estimation scheme, herein referred to as \ac{TALS} receiver. More specifically, the algorithm consists of a three-step estimation procedure that estimates one matrix at each step, while fixing the other two matrices to their values obtained at the previous estimation steps. Note that the proposed \ac{TALS} receiver is a semi-blind method, since no training sequences are required. The receiver algorithm is summarized in Algorithm \ref{PsCode_PARATUCK}. 

 The stopping criterion relies on the normalized squared error measure  computed at the end of the $i$-th iteration, given by $\epsilon_{(i)}=\sum\limits_{k=1}^K\|\mathbf{Y}[k] - \Hat{\mathbf{Y}}[k]_{(i)}\|_F^2/\|\mathbf{Y}[k]\|_F^2$, where $\Hat{\mathbf{Y}}[k]_{(i)}=\hat{\mathbf{H}}_{(i)}D_k(\mathbf{S})\hat{\mathbf{G}}_{(i)}D_k(\mathbf{W})\hat{\mathbf{X}}^{\textrm{T}}_{(i)}$. 
The convergence is declared when the difference between the reconstruction errors of two successive iterations falls below a threshold, i.e., $| \epsilon_{(i)} - \epsilon_{(i-1)}| \leq \delta$. In this work, we assume $\delta = 10^{-5}$.
The convergence criterion of \ac{TALS}  is based on the difference between the reconstruction errors computed in two successive iterations. The complexity of the \ac{TALS} receiver is dominated by the matrix inverses in steps $1$ and $3$, which have complexity orders $\mathcal{O}(TKN^2)$ and $\mathcal{O}(LKM^2)$, respectively \cite{Favier2019}. Considering the complexity of step 2, which only involves matrix products, the total complexity by iteration of the \ac{TALS} receiver is given by $\mathcal{O}(TKN^2[1 + M^2L] + KLM[NT + M])$.

\subsection{Identifiability}
The joint recovery of $\mathbf{H}$, $\mathbf{G}$, and $\mathbf{X}$ requires that the three LS problems in (\ref{func costG_PT2}), (\ref{func costH_PT2}), and (\ref{func costX_PT2}), have unique solutions, respectively. More specifically, the uniqueness of $\mathbf{H}$ requires that $\mathbf{F}$ defined in (\ref{eq:Fmatrix}) have full column-rank, which implies $TK \geq N$, while the uniqueness of $\mathbf{G}$ requires that $\left[\mathbf{\Psi}^\textrm{T} \diamond (\mathbf{X} \otimes \mathbf{H})\right]$ have full column-rank, implying $TKM \geq LN$. Likewise, the uniqueness of $\mathbf{X}$ requires that $\mathbf{E}$ defined in (\ref{eq:Ematrix}) be of full column-rank, which implies $MK \geq L$. Note that the number $K$ of transmitted blocks is the common parameter in these three conditions, which must be simultaneously satisfied. In summary, the following conditions must be simultaneously satisfied the joint uniqueness of $\mathbf{H}$, $\mathbf{G}$, and $\mathbf{X}$:
\begin{equation}
TK \geq N, \quad TKM \geq LN, \quad MK \geq L.
\label{EQ:indenti Condition}
\end{equation}
These conditions establish useful trade-offs involving the time diversities (parameters $K$ and $T$) and spatial diversities (parameters $N$, $M$, $L$) for the joint recovery of the channel and symbol matrices. More specifically, reducing the number of blocks $K$ and/or the number of symbol periods $T$ can be compensated by a corresponding increase on the number of \ac{BS} antennas $M$. As a special case, if the number of \ac{BS} antennas exceeds the number of \ac{UT} antennas ($M\geq L$), satisfying these conditions reduces to $TK\geq N$. 

Under the conditions stated above, the estimates of $\mathbf{G}$, $\mathbf{H}$, and $\mathbf{X}$ delivered by Algorithm \ref{Algorithm:TALS} are affected by scaling ambiguities that compensate each other, as follows
\begin{equation}
\Hat{\mathbf{H}}\boldsymbol{\Delta}_{H} = \mathbf{H}, \quad
\Hat{\mathbf{X}}\boldsymbol{\Delta}_{X} = \mathbf{X}, \quad
\boldsymbol{\Delta}_{H}^{-1}\Hat{\mathbf{G}}\boldsymbol{\Delta}_{X}^{-1} = \mathbf{G},\label{eq:ambiguities_paratuck}
\end{equation} 
where $\boldsymbol{\Delta}_{G}$, $\boldsymbol{\Delta}_{H}$, and $\boldsymbol{\Delta}_{X}$ are diagonal matrices. These scaling ambiguities can be handled by assuming that the first row of the symbol matrix $\mathbf{X}_{1.} \in \mathbb{C}^{1 \times L}$ contains identification symbols that are known at the receiver. Note that the columns of $\mathbf{X} \in \mathbb{C}^{T \times L}$ correspond to the $L$ data streams that are spatially multiplexed at the transmitter. Hence, the knowledge of the first row of $\mathbf{X} \in \mathbb{C}^{T \times L}$ means that the first symbol of each data stream is a known pilot. Therefore, the knowledge of $L$ pilots allows us to eliminate the scaling ambiguity by normalization. A simple choice is to assume that $\mathbf{X}_{1\bullet}=[1, 1, \ldots, 1]$, so that $\boldsymbol{\Delta}_{X}$ can be determined from the first row of the estimated symbol matrix $\Hat{\mathbf{X}}$ after convergence of Algorithm \ref{Algorithm:TALS}, and cancelled out by normalization. 

\begin{remark}
\gil{ \rev{The identifiability conditions (\ref{EQ:indenti Condition}) show that the minimum value of $K$ can be small if the block length $T$ is large. However, this minimum value may not be enough to obtain a fine \ac{CE} estimate, especially when the number of \ac{IRS} elements is large. In this case, $K$ must be increased to ensure a good performance. To overcome this practical challenge, we can resort to alternative strategies proposed in the literature, such as the partitioning of the \ac{IRS} panel into multiple sub-panels \cite{R_Zhang_Review2,Alexandropoulos_2020}. As an example, for an \ac{IRS} consisting of $N=100$ elements and partitioned into $10$ sub-panels, the minimum value of $K$ that ensures identifiability according to (\ref{EQ:indenti Condition}) is reduced by a factor of 10, for fixed values of $L$ and $M$, leading to an increased spectral efficiency.} In this case, the proposed semi-blind receiver is run independently and in parallel for each sub-panel. Moreover, if the elements within a given sub-panel are spatially correlated, one can assume the same phase shift for the entire sub-panel and further reduce $K$. Consequently, each sub-panel is associated with one aggregated channel coefficient (see, e.g. \cite{zheng_survey}) for further details on this strategy. }
\end{remark}

{\color{black} 
\section{Direct link aided Semi-Blind Receiver}
\vspace{-0ex}
\label{SEC:LOS LINK SECTION- (PARAFAC RECEIVER)}
\begin{figure}[!t]
	\centering\includegraphics[scale=0.43]{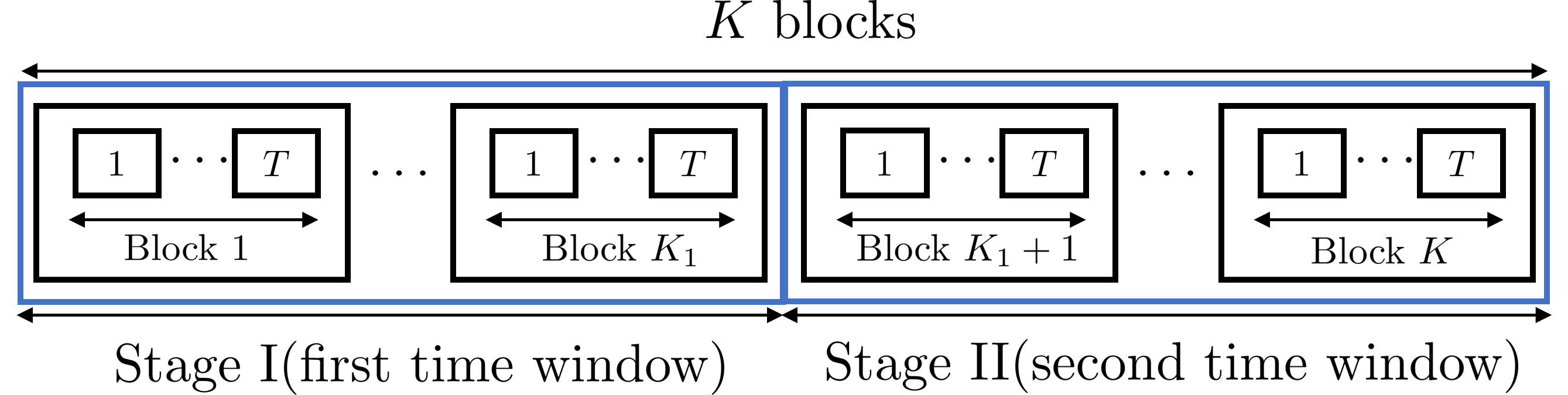}
	\caption{{\color{black}Time protocol when the assisted link via \ac{IRS} link is activated.}}
	\label{fig:Time protocol assited Link}
\end{figure}
In this section, we present an enhanced version of the semi-blind receiver derived in the previous section that exploits the direct link between \ac{UT} and \ac{BS}, whenever it is available. The idea is to dedicate part of the transmission time resources to the direct link, so that an initial estimate of the transmitted symbols and direct channel can be obtained. The initial symbol estimates are exploited as a ``warm start'' of the \ac{TALS} algorithm for estimating the \ac{UT}-\ac{IRS} and \ac{IRS}-\ac{BS} channels, while refining the estimates of the symbols and the direct channel.
To this end, we slightly modify the transmission protocol by splitting the total transmission time of $K$ blocks into two time windows of duration $T_1=K_1T$ and $T_2=K_2T$ symbol periods, respectively, where $K=K_1+K_2$ is the total number of time blocks, and $T_c=T_1+T_2$ the total transmission duration. The time protocol is depicted in Figure \ref{fig:Time protocol assited Link}. During the first time window, the UE transmits data using a {\color{black}coding} matrix $\mathbf{W}_1\in \mathbb{C}^{K_1 \times L}$, while in the second time window it uses the {\color{black}coding}  {\color{black}matrix $\mathbf{W}_2 \in \mathbb{C}^{K_2 \times L}$ and the phase shift matrix $\mathbf{S} \in \mathbb{C}^{K_2 \times N}$. A short discussion  on the design of $\mathbf{W}_2$ and $\mathbf{S}$ is provided in the Appendix \ref{Appendix B}.}

The receiver processing has two stages. In the first one, a joint estimation of the direct channel and the transmitted symbols is carried out during the first time window by exploiting the PARAFAC tensor model of the received signals. The second stage makes use of the estimated symbols in the first time window as an initialization of the \ac{TALS} algorithm that jointly estimates the involved channel matrices, while refining the symbol estimates during its iterative process. As will be shown later in our numerical experiments, the initialization of the \ac{IRS}-assisted link using the direct link estimates yields an enhanced \ac{TALS} algorithm with accelerated convergence and improved estimation accuracy.} \gil{Therefore, in stage one the \ac{IRS} is ``off", following the approach of  \cite{Zhang_Achor}, so that the received signal at \ac{BS} is given as} 
\begin{equation}
    \mathbf{Y}^{\textrm{(D)}}[k_1]= \mathbf{H}^{\textrm{(D)}}D_{k_1}(\mathbf{W}_1)\mathbf{\mathbf{X}^{\textrm{T}}} + \mathbf{B}[k_1], 
    \label{Eq:parafac_direct2}
\end{equation}
\gil{for $k_1=1, \ldots, K_1$}, where $\mathbf{W}_1 \in \mathbb{C}^{K_1 \times L}$ is the {\color{black}coding} matrix used during the first time window of $K_1$ blocks. The signal part of (\ref{Eq:parafac_direct2}) can be viewed as the $k_1$-th frontal matrix slice of a three-way tensor $\overline{\mathcal{Y}}^{\textrm{(D)}} \in \mathbb{C}^{M \times T \times K_1}$ that follows a PARAFAC decomposition {\color{black}with factor matrices $\mathbf{H}^{\textrm{(D)}}\;, \mathbf{X}\;, \mathbf{W}_1$}. \gil{By analogy with (\ref{slice_3}), the following correspondences can be deduced:
\begin{equation}
   \left( \mathbf{A}\;, \mathbf{B}\;, \mathbf{C}\right) \leftrightarrow \left(\mathbf{H}^{\textrm{(D)}}\;, \mathbf{X}\;, \mathbf{W}_1\right). 
   \label{EQ:PARAFAC_Correspondece}
\end{equation}}

From the uniqueness property of the PARAFAC model \cite{KRUSKAL197795, STEGEMAN2007540}, one can obtain a useful condition for guaranteed direct channel and symbol recovery in the general case where all the factor matrices of are unknown. In our context, however, since the {\color{black}coding} matrix $\mathbf{W}_1$ is assumed to be known at the receiver (\ac{BS}), simplified conditions can be obtained. Since $\mathbf{W}_1$ has full column-rank (which requires $K_1\geq L$), $M\geq 2$ receive antennas and $T\geq 2$ time slots are enough for the joint recovery of $\mathbf{H}^{\textrm{(D)}}$ and $\mathbf{X}$. This problem can be efficiently solved by means of the Khatri-Rao Factorization algorithm \cite{Gil_JTSP}, as will be detailed next.

{\color{black}\subsection{Stage I: Joint direct channel and symbol estimation}
\label{Sec: Estimation of the direct channel}
Starting from (\ref{Eq:parafac_direct2}), and defining
\begin{equation}
\hspace{-0.09cm}\mathbf{Y}^{\textrm{(D)}}\doteq [\textrm{vec}(\mathbf{Y}^{\textrm{(D)}}[1]), \ldots, \textrm{vec}(\mathbf{Y}^{\textrm{(D)}}[K_1])] \in \mathbb{C}^{MT \times K_1}\, ,
\label{EQ:Y^(D)}
\end{equation}
that collects the signals received during the first $K_1$ time blocks, we have
\begin{equation}
    \mathbf{Y}^{\textrm{(D)}}= (\mathbf{X} \diamond \mathbf{H}^{\textrm{(D)}})\mathbf{W}^{\textrm{T}}_1 +  \mathbf{B}\, ,
    \label{Eq:parafac_direct}
\end{equation}
where we have used property (\ref{Propertie Vec restrict}), and $\mathbf{B}=[\textrm{vec}(\mathbf{B}[1]), \ldots, \textrm{vec}(\mathbf{B}[K_1])] \in \mathbb{C}^{MT \times K_1}$ is the corresponding noise matrix. Defining
$$\mathbf{Z} \doteq (1/K_1)\mathbf{Y}^{\textrm{(D)}}\mathbf{W}^{\ast}_1= \mathbf{X} \diamond \mathbf{H}^{\textrm{(D)}} + (1/K_1)\mathbf{B}\mathbf{W}^{\ast}_1,$$ an estimate of $\mathbf{X}$ and $\mathbf{H}^{\textrm{(D)}}$ can be found using the Khatri-Rao Factorization  algorithm that solves the following problem \cite{Kibangou2009,Roemer2010}
\begin{equation}
    \min_{\mathbf{X}, \mathbf{H}^{\textrm{(D)}}}\| \mathbf{Z} -  \mathbf{X} \diamond \mathbf{H}^{\textrm{(D)}}\|_F\, ,
    \label{Eq:Cost_Hd}
\end{equation}
which is equivalent to solving $L$ rank-$1$ matrix approximation subproblems, and can be stated as
\begin{equation}
(\hat{\mathbf{X}}, \hat{\mathbf{H}}^{\textrm{(D)}})= \underset{\{\mathbf{x}_l\},\{\mathbf{h}_l^{\textrm{(D)}}\}}{\arg\min}  \sum\limits_{l=1}^L  \left\| \widetilde{\mathbf{Z}}_l - \mathbf{h}_l^{(\textrm{D})}\mathbf{x}_l^{\textrm{T}} \right\|_F\label{KRF_problem2}\, ,
\end{equation}
where $\widetilde{\mathbf{Z}}_l\doteq \textrm{unvec}_{M \times T}(\boldsymbol{\mathbf{z}}_l) \in \mathbf{C}^{M \times T}$, and $\mathbf{z}_l \in \mathbb{C}^{MT \times 1}$ denotes the $l$-th column of $\mathbf{Z}$, while $\mathbf{h}_l^{(\textrm{D})} \in \mathbf{C}^{M \times 1}$ and $\mathbf{x}_l^{\textrm{T}} \in \mathbf{C}^{1 \times T}$ are the $l$-th column of $\mathbf{H}^{\textrm{(D)}}$ and $l$-th row of $\mathbf{X}$, respectively. Due to space limitations, we have suppressed the details of the KRF algorithm. A pseudo code of this algorithm can be found in \cite{Gil_JTSP}. 

\begin{remark}
\gil{As an alternative to the KRF algorithm, one can also resort to the bilinear alternating least squares (BALS) algorithm that jointly estimates the direct channel matrix and the symbol matrix in an alternating way. In this work, we advocate using the KRF algorithm since it provides similar performance to BALS, while being a closed-form solution that affords an efficient implementation, since the $N$ involved rank-one matrix approximations can be optimized if executed in a parallel processing hardware.}
\end{remark}
\subsection{Stage II: \ac{IRS}-assisted channel estimation and symbol refinement}
After the stage I, the \ac{IRS} is turned ``on'' during the second transmission time window that spans $K_2$ blocks. In this case, the total received signal is given by the sum of the direct link and \ac{IRS}-assisted contributions, and is given by
\begin{eqnarray}
  \mathbf{Y}[k_2]&=& \mathbf{H}^{\textrm{(D)}}D_{k_2}(\mathbf{W}_2)\mathbf{X}^{\textrm{T}}\nonumber \\ &+& \mathbf{H}D_{k_2}(\mathbf{S})\mathbf{G}D_{k_2}(\mathbf{W}_2)\mathbf{X}^{\textrm{T}} + \mathbf{B}[k_2].
   \label{Signal:total} 
\end{eqnarray}
From the estimated symbol and direct channel matrices $\hat{\mathbf{X}}$ and $\hat{\mathbf{H}}^{\textrm{(D)}}$ delivered by the KRF algorithm in stage I (c.f. problem (\ref{KRF_problem2})), the interference form the direct link can be removed (or minimized) by subtracting an estimate of its contribution from the total received signal in stage II, yielding
\begin{equation}
    \mathbf{Q}[k_2] =   \mathbf{Y}[k_2] - \hat{\mathbf{H}}^{\textrm{(D)}}D_{k_2}(\mathbf{W}_2)\hat{\mathbf{X}}^{\textrm{T}}.
    \label{EQ:signal_subtracted}
\end{equation}
From (\ref{Signal:total}), we can write (\ref{EQ:signal_subtracted}) as
\begin{equation}
\mathbf{Q}[k_2] =  \mathbf{H}D_{k_2}(\mathbf{S})\mathbf{G}D_{k_2}(\mathbf{W}_2)\mathbf{X}^{\textrm{T}} + \overline{\mathbf{B}}[k_2]\, ,
\label{EQ:signal_subtracted2}
\end{equation}
where
$\overline{\mathbf{B}}[k_2]= \mathbf{B}[k_2] + \mathbf{E}_{\mathbf{H}}D_{k_2}(\mathbf{W}_2)\mathbf{E}_{\mathbf{X}}^{\textrm{T}}$ {\color{black} is the effective noise},
while $\mathbf{E_{\mathbf{H}}}\doteq \mathbf{H}^{\textrm{(D)}}- \hat{\mathbf{H}}^{\textrm{(D)}}$ and $\mathbf{E}_{\mathbf{X}}\doteq \mathbf{X}- \hat{\mathbf{X}}$ are error matrices associated with the estimates of the direct channel and symbol matrix in stage I. It is clear that the energy of the overall additive noise term in (\ref{EQ:signal_subtracted}) will depend on the energy of these error terms,  which in turn depends on the quality of the direct link compared with the \ac{IRS}-assisted link.

Note that the signal part of (\ref{EQ:signal_subtracted2}) corresponds to a PARATUCK decomposition of $\mathcal{Q} \in \mathbb{C}^{M \times T \times K_2}$, which is analogous to the (\ref{Signal:Paratuck2}), where the third mode has dimension $K_2$ (instead of $K$). Hence, estimates of the \ac{IRS}-assisted channel matrices $\mathbf{G}$ and $\mathbf{H}$, as well as refined estimates of the symbol matrix $\mathbf{X}$ can be obtained from $\mathcal{Q}$ by following the procedure discussed in Section IV.A-IV.C. This leads to a second semi-blind receiver, referred to as \ac{E-TALS}, summarized in Algorithm \ref{Algorithm:SESP}.

In addition, from the refined symbol estimates, an enhanced estimate of the direct channel matrix $\mathbf{H}^{\textrm{(D)}}$ can also be obtained at the end of stage II, i.e., at the convergence of the algorithm. More specifically, suppose that the \ac{E-TALS} algorithm has converged at the $i$-th iteration, and let $\mathbf{X}_{(i)}$ be the refined estimate of the symbol matrix obtained at the this iteration. Substituting $\Hat{\mathbf{X}}_{(i)}$ into (\ref{Eq:parafac_direct2}), a refined estimate of the direct channel can be obtained by solving the following problem
\begin{equation}
 \underset{\mathbf{H}^{(\textrm{D})}}{\textrm{min}} \,\,\sum\limits_{k_1=1}^{K_1}\big\| \mathbf{Y}^{(\textrm{D})}[k_1]  -  \mathbf{H}^{(\textrm{D})}D_{k_1}(\mathbf{W}_1)\Hat{\mathbf{X}}_{(i)}^{\textrm{T}} \big\|_F\, ,
   \label{cost_func_hd_refined}
\end{equation}
the solution of which is given by
\begin{equation}
\Hat{\mathbf{H}}^{\textrm{(D)}}= \big[\mathbf{Y}^{(\textrm{D})}[1], \ldots, \mathbf{Y}^{(\textrm{D})}[K_1]\big]\big[(\mathbf{W}_1 \diamond \Hat{\mathbf{X}}_{(i)})^{\textrm{T}}\big]^\dagger.
\end{equation}
In summary, when the direct link is available, the proposed \ac{E-TALS} receiver allows not only to improve the convergence speed of stage II by using previous symbol estimates as a ``warm start'', but also allows to continuously improve the accuracy of these symbol estimates \textit{via} the \ac{IRS}-assisted link, while enhancing the estimate of the involved channel matrices, including the direct channel matrix. As will be clear from our numerical experiments, the availability of the direct link makes \ac{E-TALS} (Algorithm 2) advantageous compared to \ac{TALS} without the availability of the direct link (Algorithm 1). 

Note that the identifiability condition discuss to TASL it is valid for \ac{E-TALS} algorithm \ref{Algorithm:SESP} with an additional restriction, which consist in $K_1 \geq L$ is required due the proposed KRF solution in first stage.

 \begin{algorithm}
 \footnotesize{
\IncMargin{1em}
	\DontPrintSemicolon
	\DontPrintSemicolon
	\SetKwData{Left}{left}\SetKwData{This}{this}\SetKwData{Up}{up}
	\SetKwFunction{Union}{Union}\SetKwFunction{FindCompress}{FindCompress}
	\SetKwInOut{Input}{input}\SetKwInOut{Output}{output}
	\textbf{Procedure}\\
	\Output{$\Hat{\mathbf{H}}$, $\Hat{\mathbf{G}}$, $\Hat{\mathbf{X}}$, and $\hat{\mathbf{H}}^{\textrm{(D)}}$}
	\BlankLine
	\Begin{    
	           $\blacksquare$ \textbf{Stage I: Joint direct channel and symbol estimation}
	           \vspace{2ex}
			   \begin{itemize}
			       \item[1.] \textit{From $\{\mathbf{Y}^{(\textrm{D})}[1], \ldots, \mathbf{Y}^{(\textrm{D})}[K_1]\}$, compute \\  $\Hat{\mathbf{H}}^{\textrm{(D)}}$ and $\Hat{\mathbf{X}}$ from the KRF algorithm}
			 \end{itemize}
			 \vspace{2ex}
			      $\blacksquare$ \textbf{Stage II: \ac{IRS}-assisted channel estimation and symbol refinement}\\
			      \vspace{2ex}
			       \Input{i = 0; \textit{initialize} $\Hat{\mathbf{X}}_{(i = 0)} = \Hat{\mathbf{X}}$,
			       }
			       \vspace{2ex}
			      \begin{itemize}
			       \item[2.] $i \leftarrow i+1$ 
			       \vspace{2ex}
			       \item[3.] From $\{\mathbf{Q}[1], \ldots, \mathbf{Q}[K_2]\}$, \textit{do}:
			        \begin{itemize}
			            \item[(a)] Compute $\Hat{\mathbf{H}}_{(i)}$ and $\Hat{\mathbf{G}}_{(i)}$ \textit{from steps 1 and 2 \\ of Algorithm \ref{Algorithm:TALS}, respectively.}
			            \item[(b)] Compute a refined symbol estimate $\Hat{\mathbf{X}}_{(i)}$ \\ \textit{from step 3 of Algorithm \ref{Algorithm:TALS}}.
			       \vspace{2ex}
			        \end{itemize}
				\item[4.] Repeat steps $2$ and $3$ until convergence.
				\vspace{2ex}
				\item[5:] From the refined estimate $\Hat{\mathbf{X}}_{(i)}$, compute a final estimate \\ of the direct link channel:
				\vspace{1ex}
				$\Hat{\mathbf{H}}^{\textrm{(D)}}= \big[\mathbf{Y}^{(\textrm{D})}[1], \ldots, \mathbf{Y}^{(\textrm{D})}[K_1]\big]\big[(\mathbf{W}_1 \diamond \Hat{\mathbf{X}}_{(i)})^{\textrm{T}}\big]^\dagger$.\\
			   \end{itemize} 
	\caption{Enhanced TALS (E-TALS)}
	\label{Algorithm:SESP}
	}
	}
\end{algorithm}

\section{Numerical Results}    
We evaluate the performance of the proposed semi-blind receivers. The \ac{CE} accuracy is evaluated in terms of the normalized mean square error (NMSE) given by
\gil{\begin{equation}
    \textrm{NMSE}(\Hat{\boldsymbol{\Pi}}) = \textcolor{black}{\frac{1}{R}\sum_{r=1}^{R} \dfrac{\|\boldsymbol{\Pi}^{(r)} - \Hat{\boldsymbol{\Pi}}^{(r)}\|_F^2}{ \|\boldsymbol{\Pi}^{(r)}\|_F^2}}\, ,
\end{equation}
where $\boldsymbol{\Pi} = \mathbf{H}, \mathbf{G}$} and $\Hat{\boldsymbol{\Pi}}^{(r)}$ denotes the estimation of the channels at the $r$-th run, and $R$ denotes the number of Monte-Carlo runs. The same definition applies to the estimated \ac{UT}-\ac{IRS} channel. We also evaluate the symbol error rate (SER) performance as a function of the signal to noise ratio (SNR) defined as  $\textrm{SNR} = 10\textrm{log}_{10}(\|\overline{\mathcal{Y}}\|_F^2/ \|\mathcal{B}\|_F^2)$,
where $\overline{\mathcal{Y}}$ is the noiseless received signal tensor generated according (\ref{eq:paratuck_scalar}), and $\mathcal{B}$ is the additive noise tensor. All the results represent an average from at least $R=3000$ Monte Carlo runs. Each run corresponds to an independent realization of the channel matrices, transmitted symbols and noise term. Regarding the channel model, we consider the Rayleigh fading case (i.e. the entries of channel matrices are independent and identically distributed zero-mean circularly-symmetric complex Gaussian random variables) as well as the geometrical channel model with a few specular paths, as described in Section II. We assume uniform linear arrays at the \ac{BS} and \ac{UT}, while the \ac{IRS} panel has a uniform rectangular array structure. The transmitted symbols follow a 16-PSK constellation.  {\color{black}When considering the direct link, we define $\alpha$ as the effective SNR gap (in dB) between direct link and the \ac{IRS}-assisted link. Otherwise stated, the average received signal power for the direct link is $\alpha$ dB smaller than that of the \ac{IRS}-assisted link. In Figures \ref{Fig: Rank-Deficient case}-\ref{Fig: SER_TALS}, we assume that the direct link is blocked and focus on the TALS receiver (Algorithm 1), while in the remaining figures the direct link is available and both \ac{TALS} and \ac{E-TALS} are considered.

\subsection{\ac{TALS} performance: NMSE, \ac{CRB} and SER}
In Figure \ref{Fig: Rank-Deficient case}, we evaluate the NMSE performance of the semi-blind \ac{TALS} receiver, while comparing it with competing \ac{CE} methods. We consider as references for comparisons the Block-LS\footnote{{\color{black}The competing \ac{CE} method of \cite{Ref_COMPII} has two stages. In the first one, the cascaded channel $\mathbf{C}_k = \mathbf{G}D_k(\mathbf{S})\mathbf{H}$ associated with each time block $k$ is individually estimated via an LS method, while in the second stage the path angles are extracted from the unstructured channel estimates. Since our semi-blind receiver in not concerned with the extraction of the angular parameters of the channel matrices (which can be done using existing methods), we compare the proposed \ac{TALS} method with the first stage of the Block-LS method of \cite{Ref_COMPII}.}} \ac{CE} method of \cite{Ref_COMPII} and the pilot-assisted PARAFAC-BALS method proposed in \cite{Gil_JTSP}. Both methods are direct competitors since they operate on the same system model as the proposed semi-blind receiver. The second is based on an iterative estimation of the \ac{UT}-\ac{IRS} and \ac{IRS}-\ac{BS} channel matrices using a BALS algorithm. However, the main difference is on the fact that these methods require the transmission of pilot sequences, while the proposed receiver jointly estimate the channel and the transmitted symbols semi-blindly. In this experiment, we assume $M=5$ antennas at the \ac{BS}, $L=2$ antennas at the \ac{UT}, and an \ac{IRS} composed of $N=64$ reflecting elements. The total transmission time consists of $K=128$ blocks of $T=5$ time slots each. \rev{The channel matrices associated with the \ac{IRS}-\ac{BS} and \ac{UT}-\ac{IRS} are generated according to a geometrical channel model, assuming a single path scenario (line of sight case). } The path directions are randomly generated according to a uniform distribution. At each Monte Carlo run, the azimuth and elevation angles are drawn within the intervals $[-\pi/2, \pi/2]$ and $[0, \pi/2]$, respectively.} 

As it can be seen from the figure, the \ac{TALS} receiver offers a more accurate overall channel estimate than Block-LS and PAFAFAC-BALS. In particular, the Block-LS method has an SNR gap of approximately $5$dB compared to \ac{TALS}. Indeed, the \ac{TALS} receiver fully exploits the trilinear structure of the received signal, and its improved performance comes from the data-aided nature of the receiver, where the symbol estimates are used to further improve the channel estimates during the iterative process. On the other hand, we should point out that the \ac{TALS} receiver is more complex than PARFAFAC-BALS and Block-LS due to the additional symbol estimation step at every iteration. 

\begin{figure}[!t]
\centering
\includegraphics[scale=0.5]{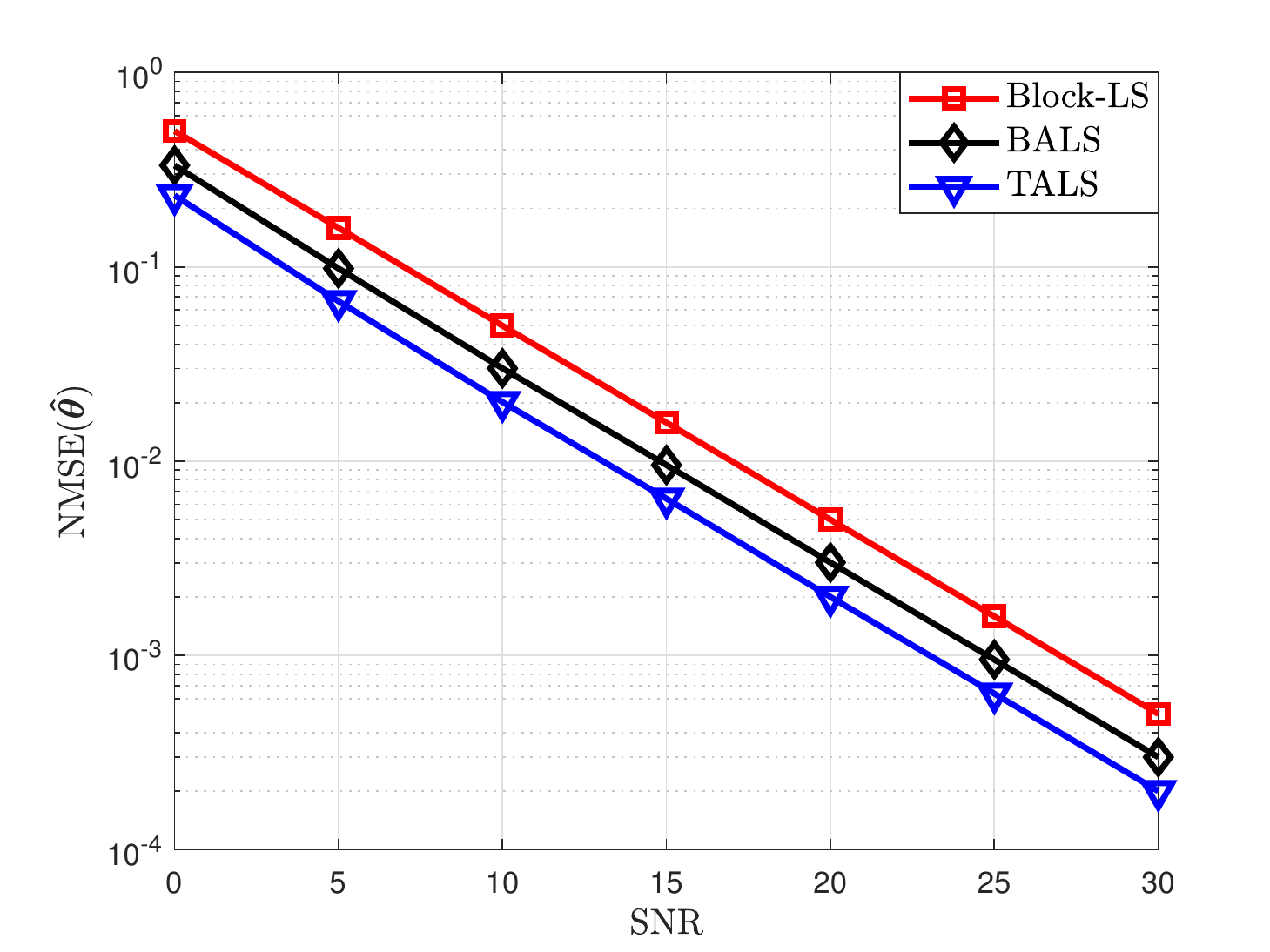}
\vspace{-3ex}
\caption{NMSE performance of \ac{TALS} in comparison with competing methods.} \label{Fig: Rank-Deficient case}
\end{figure}
\begin{figure}[!t]
\begin{center}
\includegraphics[scale=0.5]{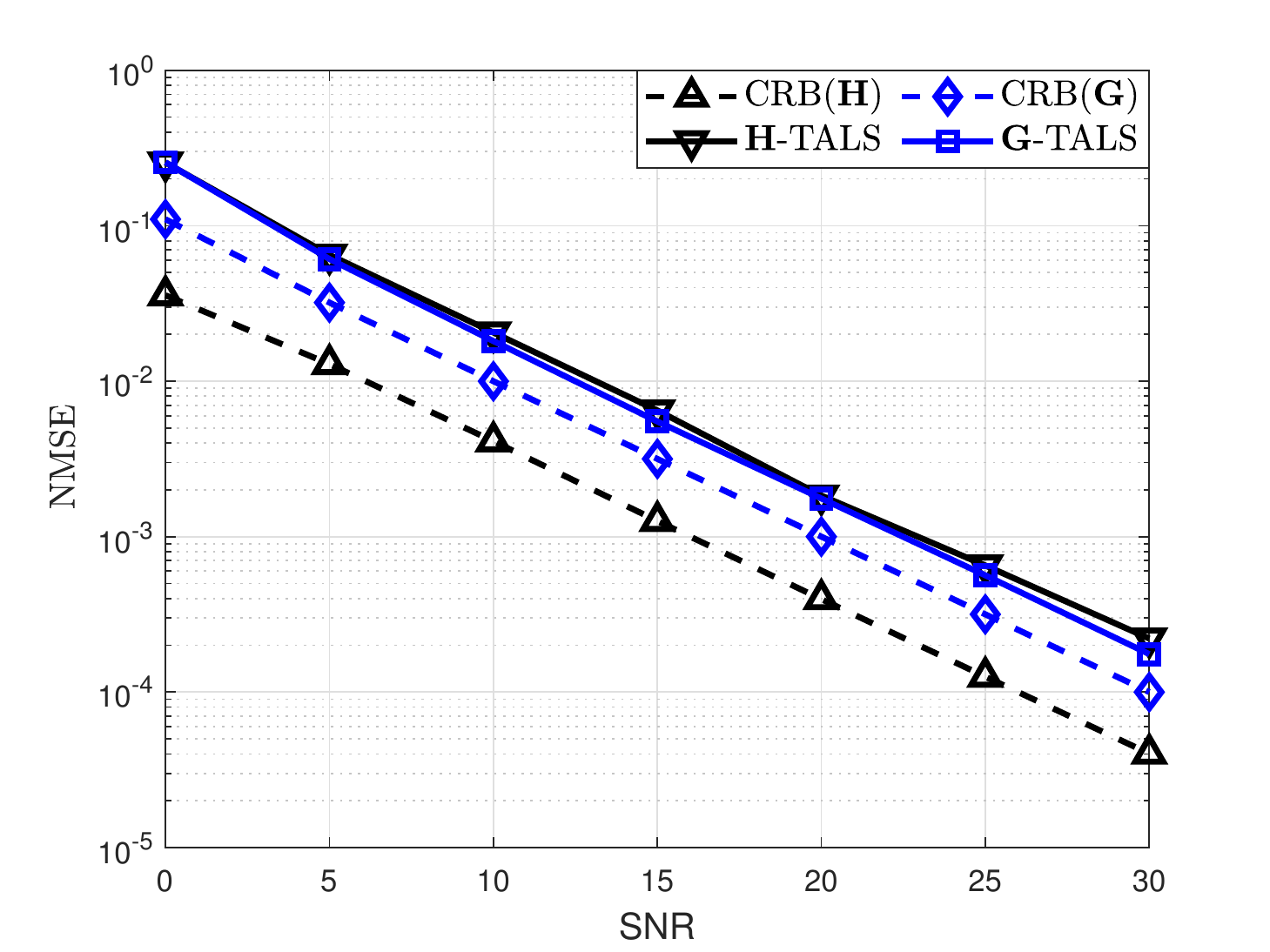}
\end{center}
	\vspace{-3ex}
\caption{Comparison with the \ac{CRB} for the estimation of $\mathbf{H}$ and $\mathbf{G}$.}
	\label{Fig: CRB_LR}
\end{figure}

In Figure \ref{Fig: CRB_LR}, we compare the NMSE performance of the individual channel matrices $\mathbf{H}$ and $\mathbf{G}$ with their corresponding \ac{CRB} references (more details given in Appendix B). Note that the NMSE curves decrease linearly with the SNR, presenting a constant gap with respect to their \ac{CRB} references regardless of the SNR value. In particular, note that the estimate of $\mathbf{G}$ is closer to its \ac{CRB} than is the estimate of $\mathbf{H}$. Figure \ref{Fig: SER_TALS} depicts the symbol error rate for some values of $N$ and $T = 2$. The other parameters follow the same setup as in Figures \ref{Fig: Rank-Deficient case} and \ref{Fig: CRB_LR}. Note that the SER performance degrades with an increasing $N$. This result is comprehensive, since more \ac{IRS} elements implies more channel coefficients to be estimated while the training time window is fixed.
\begin{figure}[!t]
\begin{center}
\includegraphics[scale=0.5]{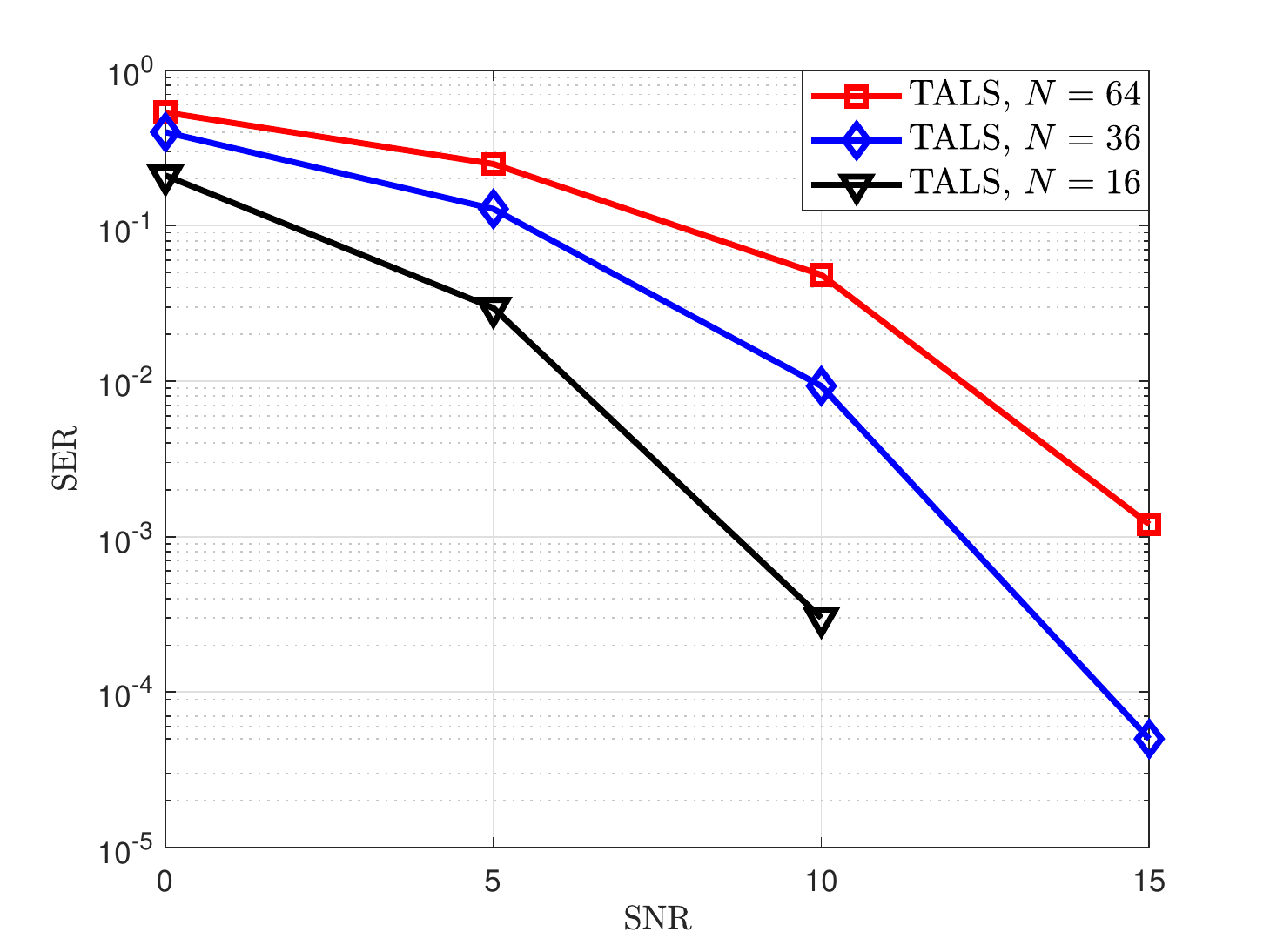}
\end{center}
	\vspace{-3ex}
\caption{SER performance of the \ac{TALS} algorithm.}
	\label{Fig: SER_TALS}
\end{figure}
{\color{black}
\begin{figure}[!t]
\begin{center}
\includegraphics[scale=0.5]{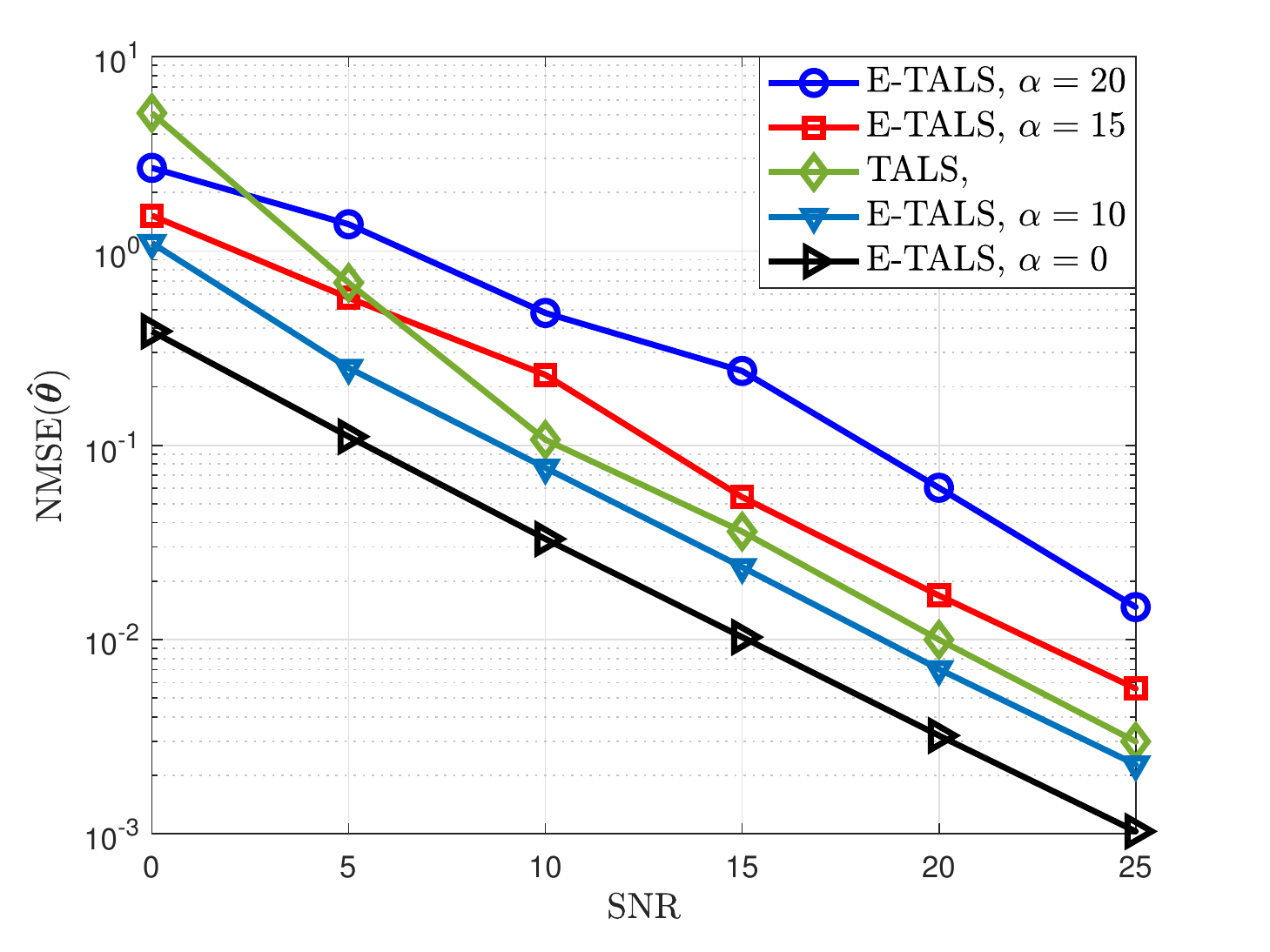}
\end{center}
	\vspace{-3ex}
\caption{Channel estimation performance  (non-orthogonal $\mathbf{\Psi}$).}
	\label{Fig: Theta_TALS x E-TALS}
\end{figure}
\gil{\subsection{\ac{E-TALS} performance: NMSE, complexity and SER}}
{\color{black} In Figure \ref{Fig: Theta_TALS x E-TALS}, we reduce the training time such that the row-orthonormal semi-unitary design of matrix $\boldsymbol{\Psi}$ is not satisfied. For this case, we consider a geometric channel model. In addition, the number of training blocks is equal to $K = 80$, each of them composed of $T = 5$ time slots. The remaining system parameters are $L = 2, \, M = 5, \, L_h = 3, \, L_g = 2$, and $3000$ Monte Carlos runs. For the \ac{E-TALS} algorithm, the splitting of the training time resources is such that $K_1 = 16$  and $K_2 = 64$. In this experiment, we consider that step $3(b)$ is ``off", which means that the estimate of the symbol matrix $\Hat{\mathbf{X}}$ delivered by the stage I of the receiver is fixed during the stage II. The results show that the presence of the direct link can be efficiently exploited by the \ac{E-TALS} receiver to further improve the accuracy of the channel estimate. In particular, we can note a performance gain of \ac{E-TALS} over \ac{TALS} for $\alpha = 0$ and $\alpha = 10$. } 
}
\begin{figure}[!t]
    \centering
    \includegraphics[scale=0.5]{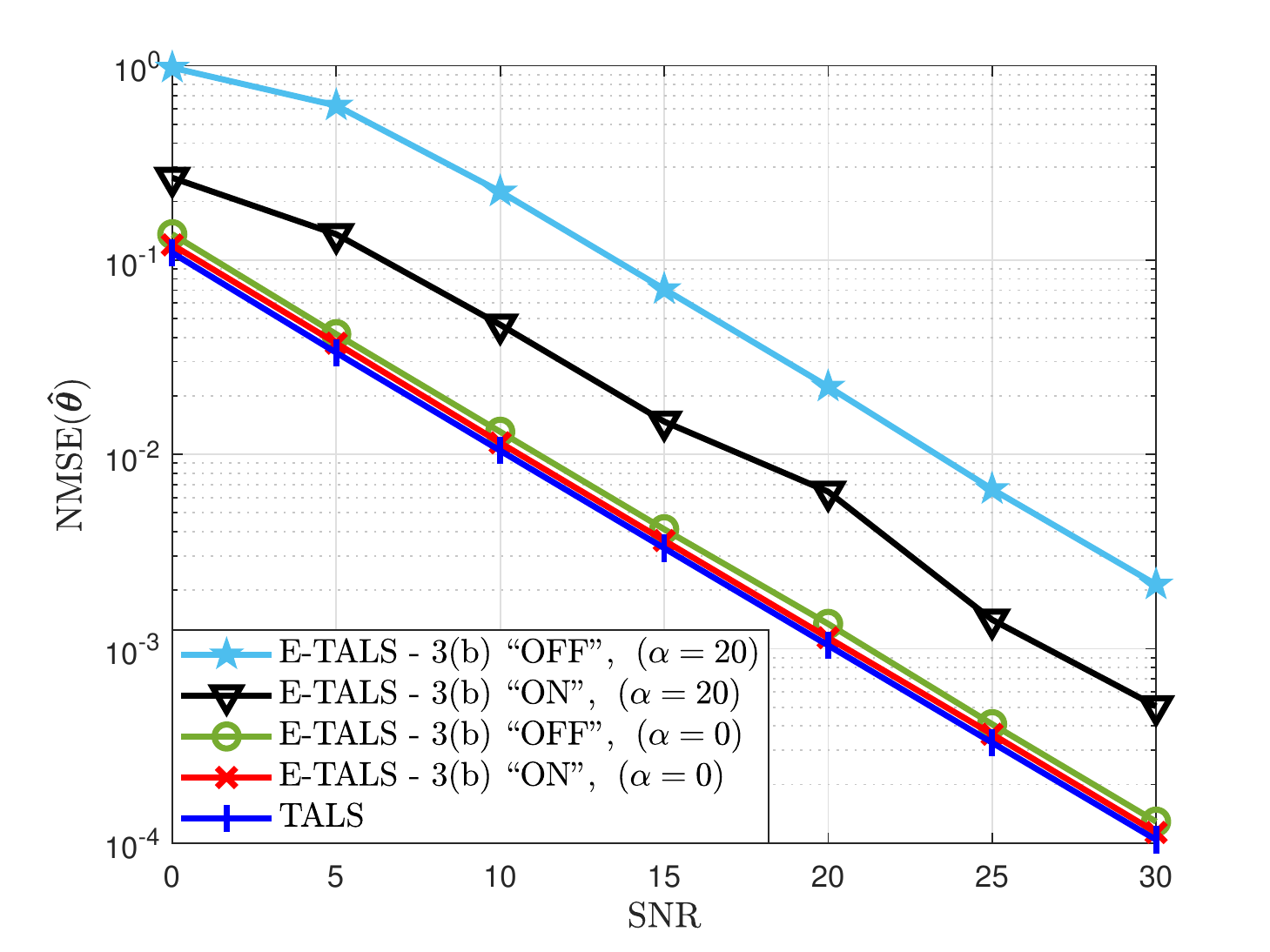}
    \vspace{-2ex}
    \caption{NMSE performance of \ac{E-TALS} for different values of $\alpha$, and the impact of symbol refinements.}
    \label{Fig:NMSE Theta for different alphas}
\end{figure}


In Figures \ref{Fig:NMSE Theta for different alphas}, \ref{Fig: Interation for different alpha} and \ref{Fig: RUNTIME}, we present how the refinement step in algorithm \ac{E-TALS} affects the \ac{CE} performance. Assuming the parameter set $\{N, M , L, T, K_1, K_2\} = \{70, 10, 2, 5, 10, 140\}$, we consider two cases: (1) the direct and the \ac{IRS}-assisted links have the same power ($\alpha = 0$), and (2) the direct link is $20$dB weaker then the \ac{IRS}- assisted link ($\alpha = 20$). Let us first consider the case 1. In this case, the result depicted in \ref{Fig:NMSE Theta for different alphas} shows that \ac{E-TALS} and \ac{TALS} present a very close NMSE performance, indicating that the impact of the refinement of the symbol estimates in the performance is negligible. Although a performance gain is not obtained, as shown in Figure \ref{Fig: Interation for different alpha}, the \ac{E-TALS} algorithm needs fewer iterations to converge in comparison to \ac{TALS}. Furthermore, we can see in Figure \ref{Fig: RUNTIME} that turning ``off" step $3$(b) of \ac{E-TALS} results in a reduced average computation time.  we can observe a significant impact on the \ac{CE}. Considering now the case 2 ($\alpha=20$dB), we can see that \ac{TALS} becomes the preferred option. This result is aligned with those of Figure \ref{Fig: Theta_TALS x E-TALS}, showing that \ac{E-TALS} is advantageous when the direct link is not too weak, as expected.

\begin{figure}[!t]
    \centering
    \vspace{-3ex}
   \includegraphics[scale=0.5]{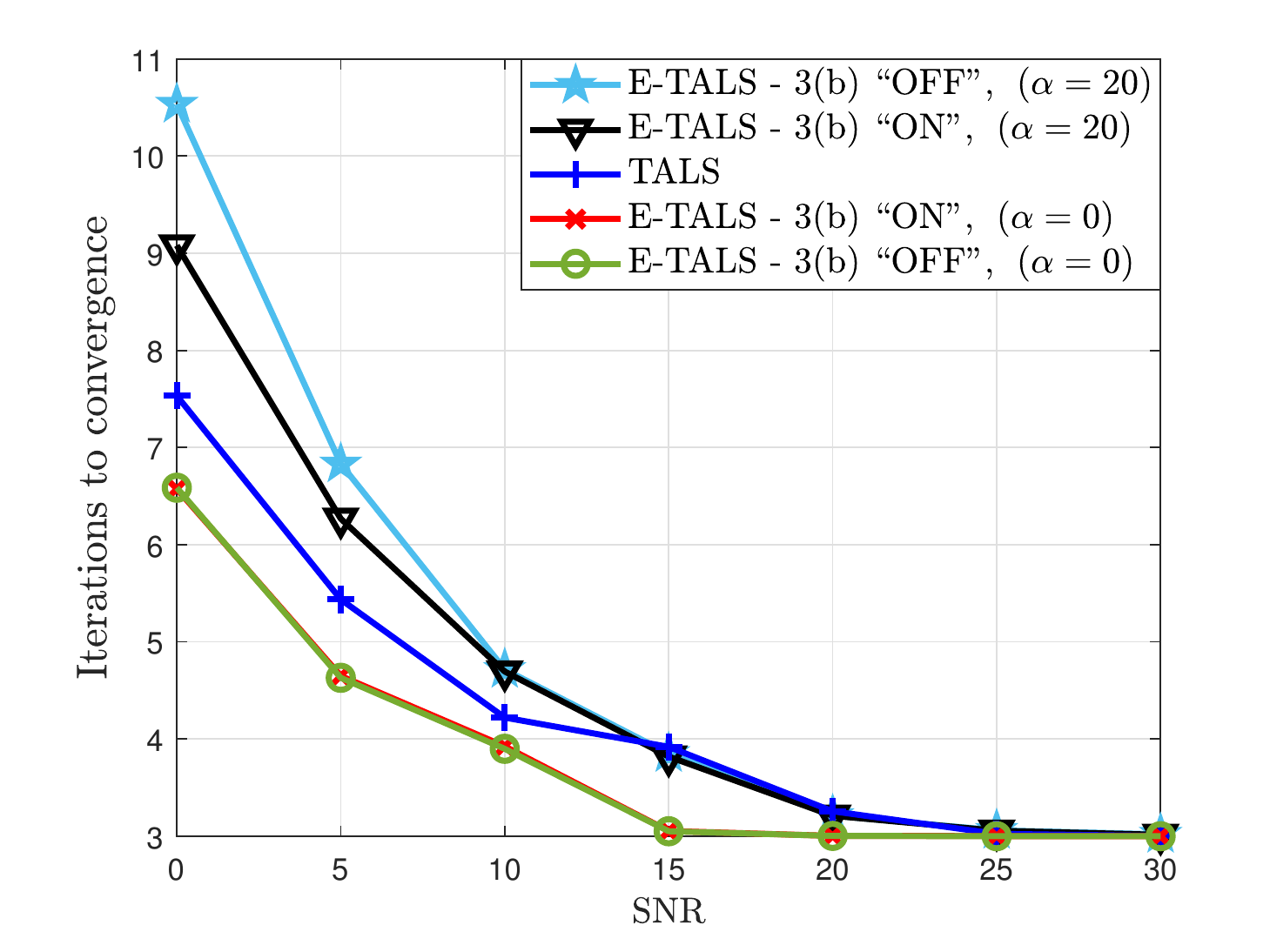}
    \caption{Convergence (number of iterations) as a function of the SNR.}
    \label{Fig: Interation for different alpha}
\end{figure} 
\begin{figure}[!t]
\centering
\vspace{-2ex}
\includegraphics[scale=0.5]{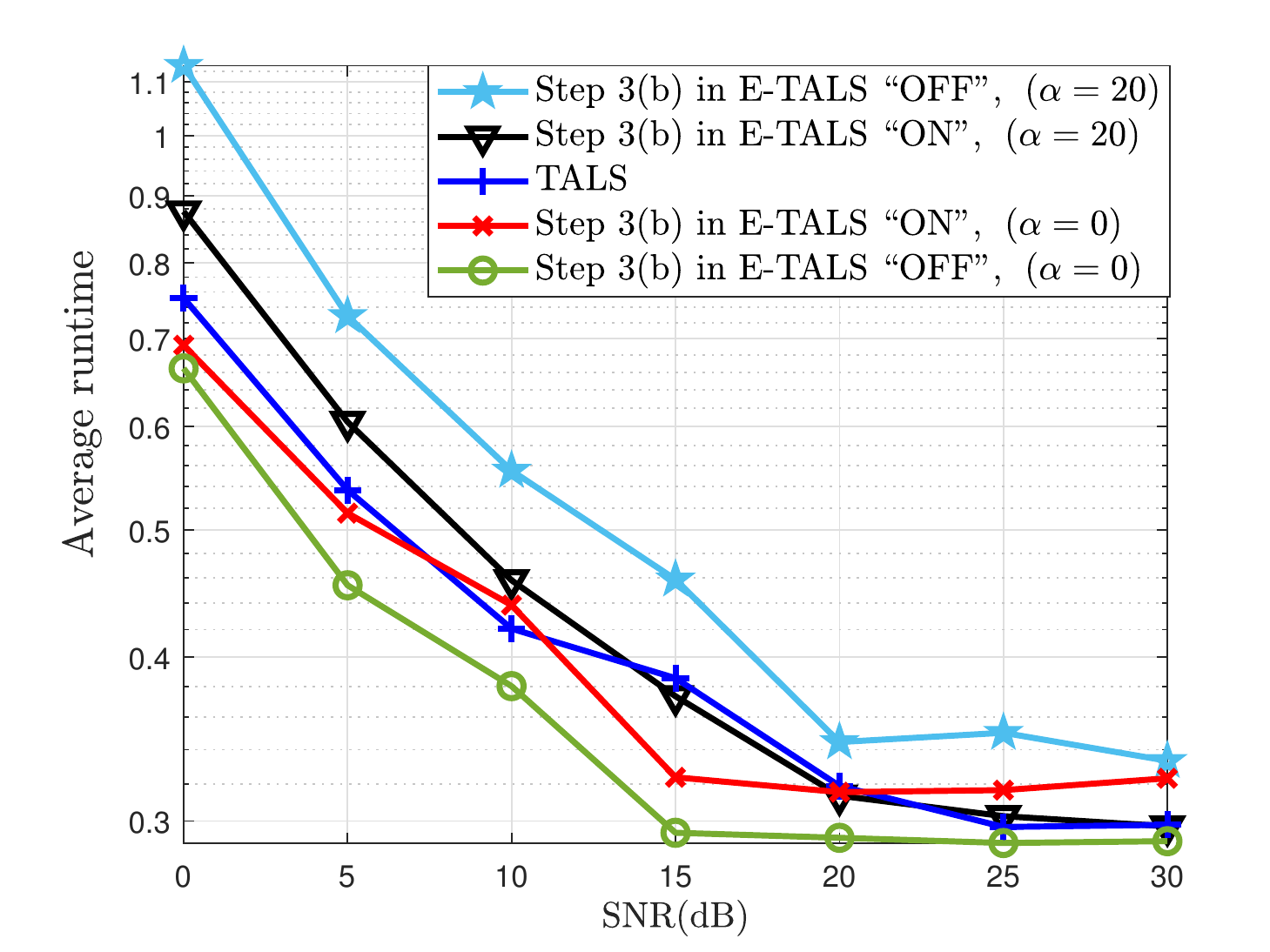}
	\vspace{-2ex}
\caption{Computation time of \ac{TALS} and \ac{E-TALS} for different settings.}
\label{Fig: RUNTIME}
\end{figure}

Figure \ref{Fig: SER for diferent N values} depicts the SER performance of \ac{E-TALS}, assuming $\alpha = 20$, $M = 10$, $T = 5$. The number of time blocks is $K = 150$, where $K_1=10$ blocks are allocated to stage I and $K_2=140$ blocks to stage II. We can clearly see that the SER associated with the refined symbol estimates provided by stage II is significantly lower than that delivered by stage I (KRF), corroborating the enhancement provided by the joint channel and symbol estimation in stage II of \ac{E-TALS}. In particular, the SER is further improved as $N$ is increased from 10 to 50. }
\begin{figure}[!t]
    \centering
    \vspace{-3ex}
    \includegraphics[scale=0.5]{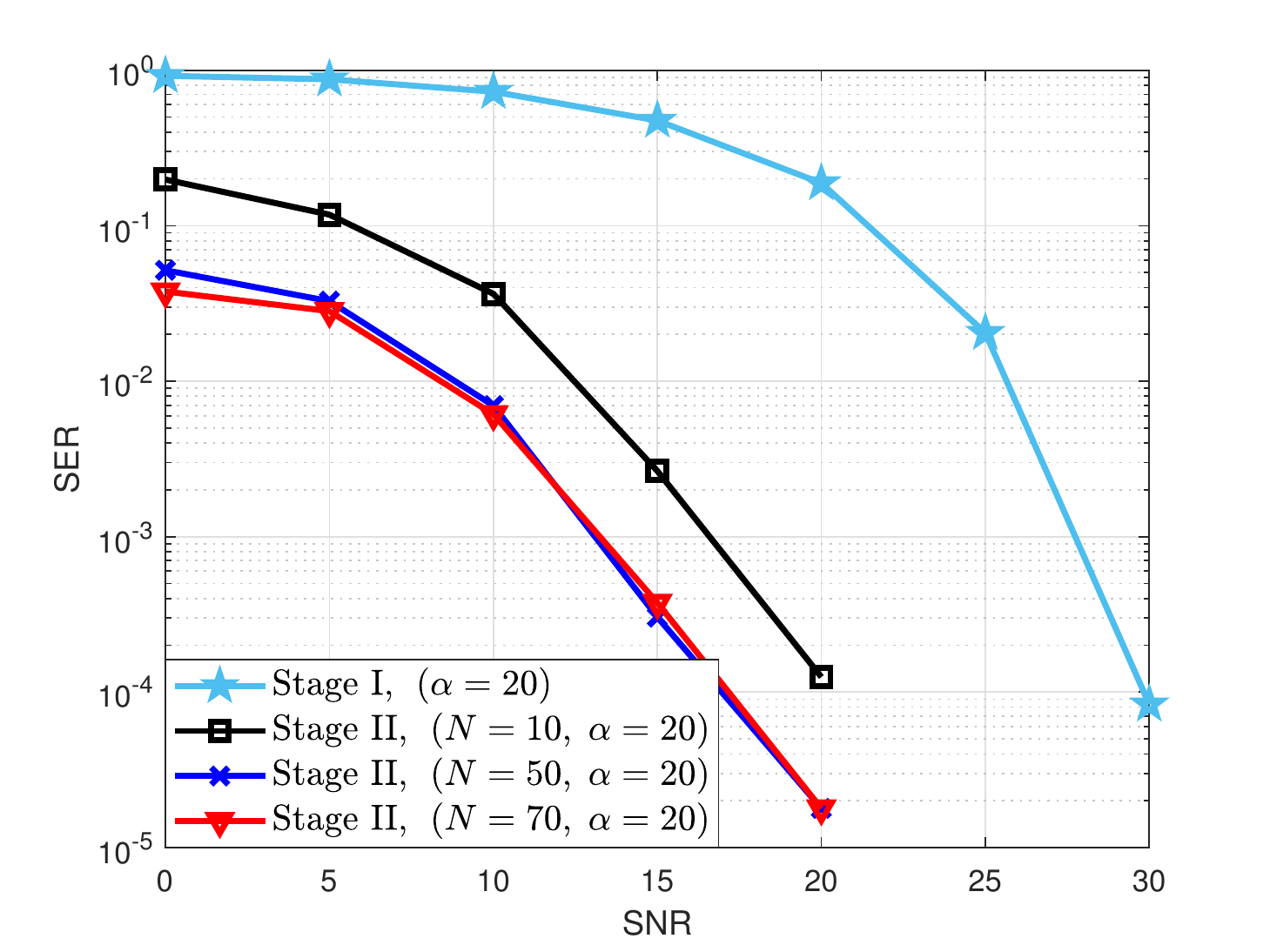}
    \vspace{-2ex}
    \caption{SER performance of \ac{E-TALS} algorithm for different values of $N$.}
    \label{Fig: SER for diferent N values}
\end{figure}
{\color{black}In Figure \ref{Fig: Hd}, we highlight the benefit of the symbol refinements provided by stage II of \ac{E-TALS} to further improve the estimate of the direct channel $\mathbf{H}^{\textrm{(d)}}$ delivered by stage I. The system parameters are $N = 50$, $M = 10$, $L = 2$, $T = 5$ and $K_1 = 10$. Recall that step $5$ of the \ac{E-TALS} algorithm makes use of the refined symbol estimates$\hat{\mathbf{X}}$ to obtain a final LS estimate of the direct channel as $\Hat{\mathbf{H}}^{\textrm{(D)}}= \big[\mathbf{Y}^{(\textrm{D})}[1], \ldots, \mathbf{Y}^{(\textrm{D})}[K_1]\big]\big[(\mathbf{W}_1 \diamond \Hat{\mathbf{X}}_{(i)})^{\textrm{T}}\big]^\dagger$. We can see that stage II of \ac{E-TALS} provides an enhanced estimate of the direct channel  compared to stage I, for both $\alpha = 0$ or $\alpha = 20$. This result confirms that the refinement of the symbol estimates obtained via the \ac{IRS}-assisted link is also beneficial to further improves the accuracy of the estimate of the direct channel, while providing estimates of the \ac{IRS}-assisted channels.}
\begin{figure}[!t]
\begin{center}
\includegraphics[scale=0.5]{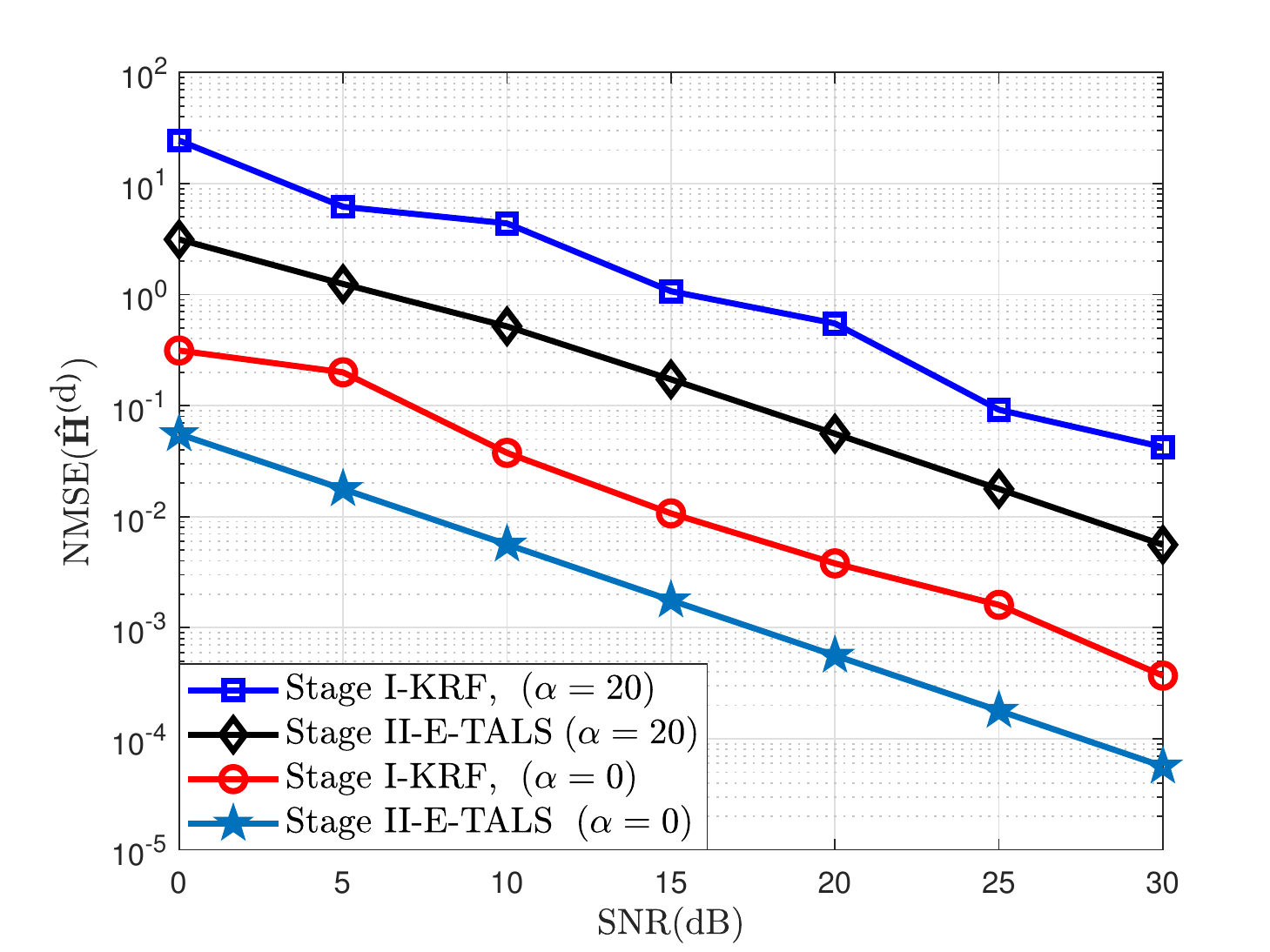}
\end{center}
	\vspace{-3ex}
\caption{NMSE of the $\mathbf{H}^{\textrm{(d)}}$ for \ac{E-TALS}.}
	\label{Fig: Hd}
\end{figure}

\begin{figure}[!t]
\hspace*{-6ex}
\centering
\vspace{-1ex} 
        \includegraphics[scale=0.55]{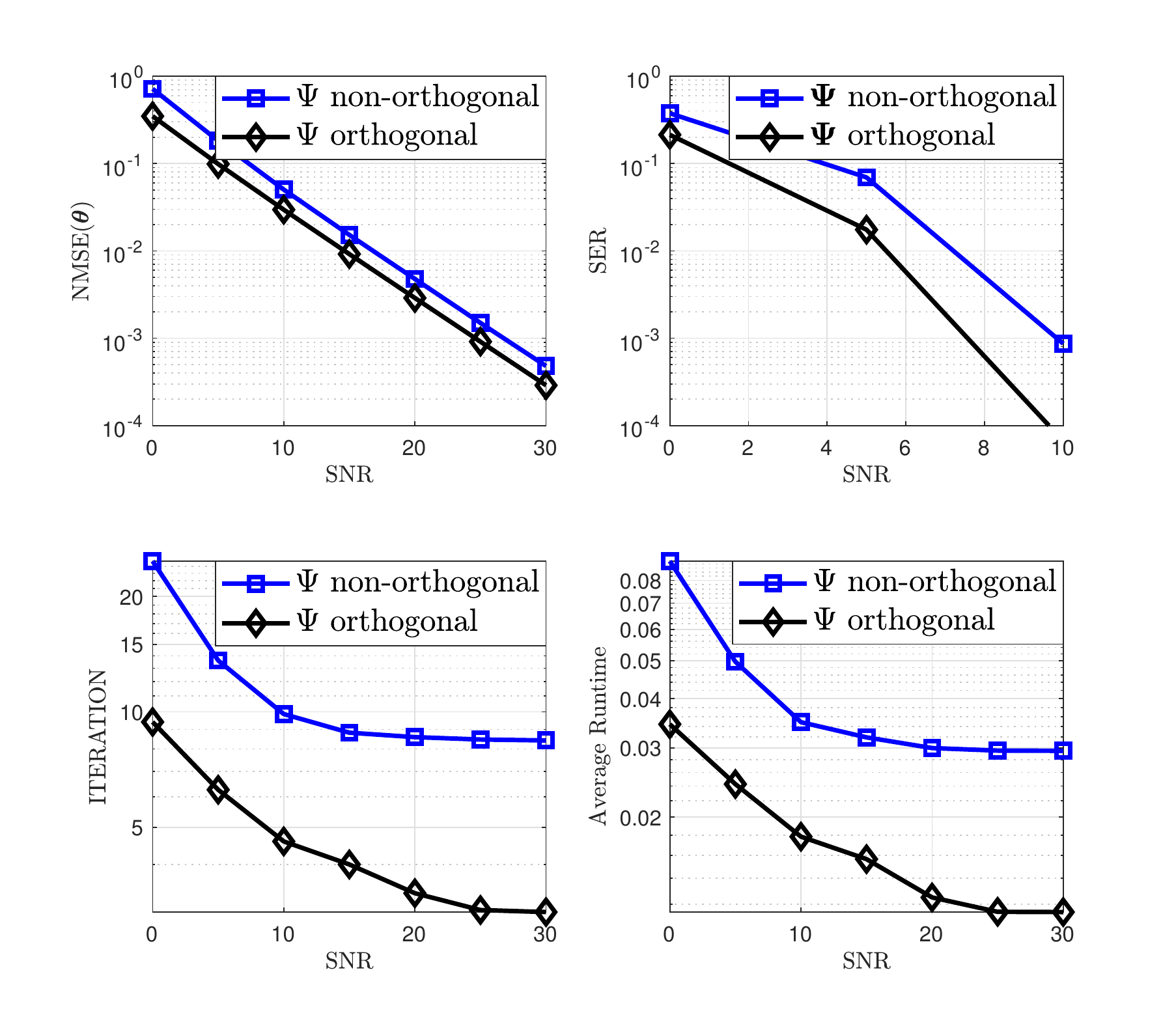}
        \caption{Orthogonal \emph{versus} non-orthogonal designs for $\boldsymbol{\Psi}$.}
        \label{Fig:Orth_vs_NoOrth}   
\end{figure}

\gil{In Figure \ref{Fig:Orth_vs_NoOrth}, \rev{we evaluate the impact of the design of $\boldsymbol{\Psi}$ on the performance of the proposed semi-blind receiver. } As it can be seen from these results, the proposed joint orthogonal design enhances the channel and symbol estimation performances. Moreover, this orthogonal design allows to reduce the overall computational cost of the semi-blind receiver, since fewer iterations are required for convergence.}

\vspace{4ex}

\section{Conclusion} 
In this paper, we have proposed a novel tensor-based semi-blind receiver design for \ac{IRS}-assisted \ac{MIMO} communication systems that exploits PARATUCK tensor model for the received signals. The proposed semi-blind receiver is a data-aided channel estimator that avoids the use of pilot sequences, while performing a joint estimation of the \ac{IRS}-\ac{BS} channel, \ac{UT}-\ac{IRS} channel, and the transmitted symbols in an iterative way by means of a \ac{TALS} (algorithm \ref{Algorithm:TALS}) {\color{black}when the direct link is unavailable or negligible or by means of \ac{E-TALS} (algorithm \ref{Algorithm:SESP}) if the direct link is available.} We have studied the design of the coding matrix and \ac{IRS} phase shift matrix, and a joint design has been proposed that optimizes the receiver performance while simplifying the \ac{CRB} derivations. Our results also indicate that the \ac{TALS} receiver yields an improved \ac{CE} accuracy than ``block-LS" and BALS algorithms, while offering a joint channel and symbol recovery, thus being a good solution for \ac{IRS}-assisted \ac{MIMO} systems, especially when pilot resources are limited or not available. 
Analytical expressions for the \ac{CRB} have been derived for the proposed semi-blind receiver. \rev{We believe that extending the proposed semi-blind receiver to the multiuser scenario, in the presence of frequency-selective channels is an important topic for future research. The generalization of the proposed approach to the multi-IRS case is also an interesting topic for future work.}

\begin{appendices}

\section{Expected Cram\'er Rao lower Bound}
\label{Appendix A}
{\color{black} The \ac{CRB} is the lowest estimation accuracy that an unbiased estimator can reach. If $\boldsymbol{\Hat{\theta}}$ is an unbiased estimate of $\boldsymbol{\theta}$, the MSE measurements is lower bounded by the \ac{CRB} such as,
\begin{equation}
    \mathbb{E}\|\boldsymbol{\theta} - \boldsymbol{\Hat{\theta}}\|^2 \geq Tr\{\textrm{CRB}{\boldsymbol{(\theta)}}\},
\end{equation}
where \ac{CRB}($\boldsymbol{\theta}$) is given as the inverse of the Fisher Information Matrix (FIM), denoted by $\mathbf{F}(\boldsymbol{\theta})$, such as $\textrm{CRB}(\boldsymbol{\theta}) = \mathbf{F}(\boldsymbol{\theta})^{-1}$.
An extension for complex-valued parameters can be as in \cite{Favier2019}, making by structured parameters vector $\boldsymbol{\theta}_c = \left[ \boldsymbol{\bar{\theta}}^{\textrm{T}} \boldsymbol{\widetilde{\theta}}^{\textrm{T}} \right]^{\textrm{T}}$, where $\boldsymbol{\bar{\theta}}=Re(\boldsymbol{\theta})$, and $\boldsymbol{\widetilde{\theta}}=Im(\boldsymbol{\theta})$. Thereby, with a nuisance parameter $\gamma$ the expected \ac{CRB} for complex-valued random parameters is given as
\begin{equation}\label{CRB_tot}
    \mathbb{E}\|\boldsymbol{\theta}_c - \boldsymbol{\Hat{\theta}}_c\|^2 \geq \mathbb{E}_{\boldsymbol{\bar{\theta}},\boldsymbol{\widetilde{\theta}},\gamma}\left\{Tr\{\textrm{CRB}\boldsymbol{(\bar{\theta})}\} + Tr\{\textrm{CRB}\boldsymbol{(\widetilde{\theta})}\}\right\}.
\end{equation}

For an observation vector that follows a complex circular Gaussian distribution, $\mathbf{y} \sim CN(\boldsymbol{\mu},\mathbf{R})$, a useful formula, used to obtain the FIM, is the Slepian-Bangs (SB) Formula \cite{P_Stoica}:
\begin{eqnarray}
\left[ \mathbf{F}(\boldsymbol{\theta}) \right]_{i,j} &=& 2Re\left\{ \left( \dfrac{\partial \boldsymbol{\mu}}{\partial\boldsymbol{[\theta_c}]_i}\right)^{\textrm{H}} \mathbf{R}^{-1}\left( \dfrac{\partial \boldsymbol{\mu}}{\partial\boldsymbol{[\theta_c}]_j}\right) \right \}
\label{SB_Formula_PART1}\\
& + & Tr \left \{\left( \dfrac{\partial \mathbf{R}}{\partial\boldsymbol{[\theta_c}]_i}\right)\mathbf{R}^{-1} \left( \dfrac{\partial \mathbf{R}}{\partial\boldsymbol{[\theta_c}]_j}\right)\mathbf{R}^{-1} \right \}.
\label{SB_Formula_PART2}
\end{eqnarray}
such that,
\begin{equation}
    \mathbf{F}(\boldsymbol{\theta}_c) = 2\left[  
     \begin{array}{lr}
          \overline{\mathbf{M}}  &  -\widetilde{\mathbf{M}}\\
           \widetilde{\mathbf{M}} & \overline{\mathbf{M}} 
     \end{array}
    \right],
    \label{Fim_matrix}
\end{equation}
and by deriving analytically $\mathbf{F}(\boldsymbol{\theta}_c)^{-1}$ using the schur complement  method and considering the trace operator we obtain
\begin{equation}
Tr\{\textrm{CRB}(\bar{\boldsymbol{\theta}})\} = \dfrac{1}{2}Tr\left\{ \left( \overline{\mathbf{M}} + \widetilde{\mathbf{M}}\overline{\mathbf{M}}^{-1}\widetilde{\mathbf{M}} \right)^{-1} \right\},
\label{CRB_Real_PART}
\end{equation}
\begin{equation}
\begin{split}
Tr\{\textrm{CRB}(\widetilde{\boldsymbol{\theta}})\}  =   \dfrac{1}{2} & Tr \left\{\overline{\mathbf{M}}^{-1} - \overline{\mathbf{M}}^{-1}\widetilde{\mathbf{M}}\left( \overline{\mathbf{M}}\right. \right. + \\
& \left. \left. +\widetilde{\mathbf{M}}\overline{\mathbf{M}}^{-1}\widetilde{\mathbf{M}}\right)^{-1}\widetilde{\mathbf{M}}\overline{\mathbf{M}}^{-1} \right\}.
\end{split}
\label{CRB_IM_PART}
\end{equation}
The matrices $\overline{\mathbf{M}}$ and $\widetilde{\mathbf{M}}$ are defined in posterior subsections according to parameters of observation.
}

The \ac{CRB} derivations for proposed semi-blind receiver are split into two parts. In the first part, we derive the \ac{CRB} for the \ac{IRS}-\ac{BS} channel $\mathbf{H}$, whereas in the second part the \ac{CRB} for the \ac{UT}-\ac{IRS} channel $\mathbf{G}$ is derived.

\subsection{\ac{CRB} for the \ac{UT}-\ac{IRS} channel}
Here, the \ac{UT}-\ac{IRS} channel $\mathbf{G}$ is viewed as unknown nuisance, and the \ac{CRB} is derived for the \ac{IRS}-\ac{BS} channel $\mathbf{H}$.
Let
\begin{eqnarray}
&\boldsymbol{\theta}_c &= [\bar{\mathbf{g}}^{\textrm{T}}\ \ \widetilde{\mathbf{g}}^{\textrm{T}}]^{\textrm{T}}, \quad \mathbf{g} = \textrm{vec}(\mathbf{G})\\
&\boldsymbol{\gamma} &= [\bar{\mathbf{h}}^{\textrm{T}}\ \ \widetilde{\mathbf{h}}^{\textrm{T}}\ \ \textrm{vec}(\mathbf{X})^{\textrm{T}}]^{\textrm{T}},
\end{eqnarray}
where, 
\begin{equation}
\boldsymbol{\mu}_2 = \mathbf{C}\mathbf{g}\quad \textrm{and} \quad \mathbf{R}_2 = \sigma^2\mathbf{I}_{TKM}.
\end{equation}
The \ac{CRB} for $\mathbf{G}$ is given by summing (\ref{CRB_Real_PART}) and (\ref{CRB_IM_PART}) where $\overline{\mathbf{M}} = Re\{\mathbf{C}^{\textrm{H}}\mathbf{R}^{-1}\mathbf{C}\}$ and $\widetilde{\mathbf{M}} = Im\{\mathbf{C}^{\textrm{H}}\mathbf{R}^{-1}\mathbf{C}\}$, where $\mathbf{C}^{\textrm{H}}\mathbf{R}^{-1}\mathbf{C}$ can be expanded as follows
\begin{equation}
\begin{aligned}
\mathbf{C}^{\textrm{H}}\mathbf{R}^{-1}\mathbf{C} &= (1/\sigma^2)\left(\mathbf{\Psi}^{\textrm{T}} \diamond (\mathbf{X} \otimes \mathbf{H}) \right)^{\textrm{H}}\left(\mathbf{\Psi}^{\textrm{T}} \diamond (\mathbf{X} \otimes \mathbf{H}) \right)\\
&=  (1/\sigma^2)\left( \mathbf{\Psi}^{\ast}\mathbf{\Psi}^{\textrm{T}} \odot (\mathbf{X} \otimes \mathbf{H})^{\textrm{H}}(\mathbf{X} \otimes \mathbf{H}) \right)\\
&=  (1/\sigma^2)\left( \mathbf{\Psi}^{\ast}\mathbf{\Psi}^{\textrm{T}} \odot \left( \mathbf{X}^{\textrm{H}}\mathbf{X} \otimes \mathbf{H}^{\textrm{H}}\mathbf{H} \right)\right).
\end{aligned}
\label{Analitic PSI}
\end{equation}

Under the assumption that $\mathbf{\Psi}$ is a semi-unitary matrix satisfying $\mathbf{\Psi}^{\ast}\mathbf{\Psi}^{\textrm{T}}= K\mathbf{I}_{LN}$ (see our discussion in Appendix \ref{Appendix B}), we have
\begin{equation}
\begin{split}
\mathbf{C}^{\textrm{H}}\mathbf{R}^{-1}\mathbf{C} &= (K/\sigma^2)\left( \mathbf{I}_{LN} \odot \left( \mathbf{X}^{\textrm{H}}\mathbf{X} \otimes \mathbf{H}^{\textrm{H}}\mathbf{H} \right)\right).
\end{split}
\label{Analitic PSI FINAL}
\end{equation}

Note that according to (\ref{Analitic PSI FINAL}), $\mathbf{C}^{\textrm{H}}\mathbf{R}^{-1}\mathbf{C}$ is a real-valued diagonal matrix, which implies $\widetilde{\mathbf{M}} = \mathbf{0}$. As a consequence (\ref{Fim_matrix}) is block diagonal matrix, meaning that the real and imaginary parts are decoupled. Plugging (\ref{Analitic PSI FINAL}) into (\ref{CRB_Real_PART}) and (\ref{CRB_IM_PART}), the \ac{CRB} for $\mathbf{G}$ is obtained.

\subsection{\ac{CRB} for the \ac{IRS}-\ac{BS} channel}
Here, the \ac{IRS}-BS channel $\mathbf{H}$ is treated as an unknown nuisance and the \ac{CRB} is derived for the \ac{UT}-\ac{IRS} channel $\mathbf{G}$. Thus 
\begin{eqnarray}
&\boldsymbol{\theta}_c &= [\bar{\mathbf{h}}^{\textrm{T}}\ \widetilde{\mathbf{h}}^{\textrm{T}}]^{\textrm{T}}, \quad \mathbf{h} = \textrm{vec}(\mathbf{H})\\
&\boldsymbol{\gamma} &= [\bar{\mathbf{g}}^{\textrm{T}}\ \widetilde{\mathbf{g}}^{\textrm{T}}\ \textrm{vec}(\mathbf{X})^{\textrm{T}}]^{\textrm{T}}.
\end{eqnarray}

Applying the vec(.) operator to (\ref{Estimate_G}), we obtain the following noisy observation vector
\begin{equation}
    \mathbf{y}_1 = \left( \mathbf{F} \otimes \mathbf{I}_{M} \right)\mathbf{h}
   = \mathbf{P}\mathbf{h},
\end{equation}
where, $\mathbf{y}_1 = \textrm{vec}(\mathbf{Y}_1)$ and $\mathbf{y}_1 \sim CN\left( \boldsymbol{\mu}_3, \mathbf{R}_3\right)$, and
\begin{eqnarray}
\boldsymbol{\mu}_3 &=& \mathbf{P}\mathbf{h}\\
\mathbf{R}_3& = & \sigma^2\mathbf{I}. 
\end{eqnarray}
We have $\overline{\mathbf{M}} = Re\{ \mathbf{P}^{\textrm{H}}\mathbf{R}_2^{-1}\mathbf{P}\}$ and $\widetilde{\mathbf{M}} = Im\{ \mathbf{P}^{\textrm{H}}\mathbf{R}_2^{-1}\mathbf{P}\}$.
Then,
\begin{equation}
    \begin{split}
      \mathbf{P}^{\textrm{H}}\mathbf{R}_{2}^{-1}\mathbf{P} &=(1/\sigma^2)\left( \mathbf{F} \otimes \mathbf{I}_{M} \right)^{\textrm{H}} \left( \mathbf{F} \otimes \mathbf{I}_{M} \right)\\
      & = (1/\sigma^2)\left( \mathbf{F}^{\textrm{H}}\mathbf{F} \otimes \mathbf{I}_{M} \right)^{\textrm{H}}.
    \end{split}
\end{equation}
Finally, from the Slepian-Bangs formula, 
the \ac{CRB} for $\mathbf{H}$ is obtained by summing (\ref{CRB_Real_PART}) and (\ref{CRB_IM_PART}).


\section{Design of $\mathbf{W}$ and $\mathbf{S}$ and its implications}
\label{Appendix B}
In this appendix, we discussed the design of the {\color{black}coding} matrix $\mathbf{W}$ and the \ac{IRS} phase shift matrix $\mathbf{S}$ from their Khatri-Rao product combination
$\mathbf{\Psi}=\mathbf{W} \diamond \mathbf{S} \in \mathbb{C}^{LN \times K}$. 
{\color{black}Assuming that $\mathbf{\Psi}$ is a Vandermonde matrix constructed by truncating a $K \times K$ DFT matrix to its first $LN$ rows, with $LN \leq K$, $\mathbf{W}$ and $\mathbf{S}$ given by (\ref{contruct W}) and (\ref{Construct S}) can be obtained from an exact Khatri-Rao factorization of $\mathbf{\Psi}$. Let us consider that $\mathbf{\Psi} \in \mathbb{C}^{LN \times K}$, with $LN \leq K$, is a Vandermonde matrix constructed by truncating a $K \times K$ matrix to its first $LN$ rows. Defining $\psi_{k}\doteq e^{-j2\pi (k-1)/K}$, $k=1, \ldots, K$, as the generators of $\mathbf{\Psi}$ yields
\begin{equation}
    \mathbf{\Psi} = \left[ 
    \begin{array}{cccc}
     1    & 1 & \dots & 1 \\ 
     \psi_1   & \psi_2 &  \dots & \psi_{K}\\
     \vdots & \vdots &  \dots & \vdots\\
     \psi^{(LN-1)}_1  & \psi^{(LN-1)}_2 & \dots & \psi^{(LN-1)}_{K}\\
    \end{array}
    \right]\; .
    \label{Design PSI}
\end{equation}
Since the $k$-th column of $\mathbf{\Psi}$ is a Vandermonde vector, it can be factorized as Kronecker product of two Vandermonde vectors with generators $\psi^N_k$ and $\psi^k$, respectively, as follows
\begin{equation}
    \left[ 
    \begin{array}{c}
        1 \\
        \vdots\\
        \psi^{(LN-1)}_k
    \end{array}
    \right] \; = 
    \left[ 
    \begin{array}{c}
        1 \\
        \vdots\\
        \psi^{N(L-1)}_k
    \end{array}
    \right]\; \otimes 
    \left[ 
    \begin{array}{c}
        1 \\
        \vdots\\
        \psi^{(N-1)}_k
    \end{array}
    \right].
\end{equation}
Defining $\mathbf{W}_{k \bullet}\doteq \left[1, \psi^N_k, \ldots, \psi^{N(L-1)}_k\right] \in \mathbb{C}^{1 \times L}$ and $\mathbf{S}_{k \bullet}\doteq \left[1, \psi_k, \ldots, \psi^{(N-1)}_k\right] \in \mathbb{C}^{1\times N}$, we have
\begin{equation}
\mathbf{\Psi}_{\bullet k}= \mathbf{W}^{\textrm{T}}_{k \bullet} \otimes \mathbf{S}^{\textrm{T}}_{k \bullet}, \quad k=1, \ldots, K,
\end{equation}
or, equivalently,
\begin{equation}
\begin{split}
\mathbf{\Psi}&=[\mathbf{\Psi}_{\bullet 1}, \ldots, \mathbf{\Psi}_{\bullet K}]\\
&= [\mathbf{W}^{\textrm{T}}_{1 \bullet} \otimes \mathbf{S}^{\textrm{T}}_{1 \bullet}, \ldots,  \mathbf{W}^{\textrm{T}}_{K \bullet} \otimes \mathbf{S}^{\textrm{T}}_{K\bullet}]\\
&=\mathbf{W}^{\textrm{T}} \diamond \mathbf{S}^{\textrm{T}} \in \mathbb{C}^{LN \times K},
\end{split}
\end{equation}
where
\begin{equation}
    \mathbf{W} = \left[  
    \begin{array}{c} \mathbf{W}_{1 \bullet} \\ 
    \vdots \\\mathbf{W}_{K \bullet} \end{array}\right]= \left[  
    \begin{array}{cccc}
      1  & \psi^{N}_1 & \dots & \psi^{N(L-1)}_1\\
      \vdots & \vdots & \ddots & \vdots\\
      1 & \psi^{N}_{K} & \dots & \psi^{N(L-1)}_{K}
    \end{array}
    \right]\; \in \mathbb{C}^{K \times L},
    \label{contruct W}
\end{equation}
and
\begin{equation}
    \mathbf{S} = \left[  
    \begin{array}{c} \mathbf{S}_{1 \bullet} \\ 
    \vdots \\\mathbf{S}_{K \bullet} \end{array}\right]= \left[  
    \begin{array}{cccc}
      1  &  \psi_1 & \dots & \psi^{(N-1)}_1\\
      \vdots & \vdots & \ddots & \vdots\\
      1 & \psi_{K} & \dots & \psi^{(N-1)}_{K}
    \end{array}
    \right]\; \in \mathbb{C}^{K \times N}.
    \label{Construct S}
\end{equation}}
In order to show that the assumption $\mathbf{\Psi}^{\ast}\mathbf{\Psi}^{\textrm{T}}=K\mathbf{I}_{LN}$ implies the equivalence between (\ref{H_hat}) and (\ref{H_hat_2}), recall the LS estimation step for $\mathbf{G}$ given by (\ref{H_hat}), which involves computing the left pseudo-inverse of matrix $\mathbf{C} = \left[\mathbf{\Psi}^\textrm{T} \diamond (\mathbf{X} \otimes \mathbf{H})\right]$. Taking the Khatri-Rao structure of $\mathbf{C}$ into account, and using property (\ref{Propertie Hadmard x Khatri}), we have
\begin{equation}
 \begin{aligned}
\hspace{-0.27cm}\mathbf{C}^{\dagger}&=(\mathbf{C}^{\textrm{H}}\mathbf{C})^{-1}\mathbf{C}^{\textrm{H}}\\
&=\left( \mathbf{\Psi}^{\ast}\mathbf{\Psi}^{\textrm{T}} \odot \left( \mathbf{X}^{\textrm{H}}\mathbf{X} \otimes \mathbf{H}^{\textrm{H}}\mathbf{H} \right)\right)^{-1}\left( \mathbf{\Psi} \diamond \left(\mathbf{X} \otimes \mathbf{H} \right)\right)^{\textrm{H}}\\
&=\left( K\mathbf{I}_{LN} \odot \left( \mathbf{X}^{\textrm{H}}\mathbf{X} \otimes \mathbf{H}^{\textrm{H}}\mathbf{H} \right)\right)^{-1}\left( \mathbf{\Psi} \diamond \left(\mathbf{X} \otimes \mathbf{H} \right)\right)^{\textrm{H}}.
\end{aligned}
\label{Step two TALS}  
\end{equation}
Defining $\mathbf{Q} = \mathbf{X} \otimes \mathbf{H} \in \mathbb{C}^{TM \times LN}$, and using the property $(\mathbf{A}\otimes \mathbf{B})(\mathbf{C}\otimes \mathbf{D})= (\mathbf{A}\mathbf{C}\otimes \mathbf{B}\mathbf{D})$, equation (\ref{Step two TALS}) can be rewritten as 
\begin{equation}
\mathbf{C}^{\dagger} = (1/K)\left(\mathbf{I}_{LN} \odot (\mathbf{Q}^{\textrm{H}}\mathbf{Q})\right)^{-1}\left( \mathbf{\Psi} \diamond \mathbf{Q}\right)^{\textrm{H}}.
\label{Step two TALS Final0} 
\end{equation}
Since the Hadamard product in (\ref{Step two TALS Final0}) will null out the non-diagonal terms of the Gramian $\mathbf{Q}^{\textrm{H}}\mathbf{Q}$, this equation can be simplified to
\begin{equation}
\mathbf{C}^{\dagger} = (1/K)\boldsymbol{\Sigma}^{-1}_{\mathbf{Q}}(\mathbf{\Psi} \diamond \mathbf{Q})^{\textrm{H}},
\label{Step two TALS Final} 
\end{equation}
where $\boldsymbol{\Sigma}_{\mathbf{Q}}$ is given in (\ref{EQ:Sigma_Q}).
\end{appendices}



\renewcommand\baselinestretch{.97}

\end{document}